\documentclass[review,twoside,web]{ieeecolor}

\usepackage{etoolbox}
\makeatletter
\@ifundefined{color@begingroup}%
{\let\color@begingroup\relax
\let\color@endgroup\relax}{}%
\def\fix@ieeecolor@hbox#1{%
\hbox{\color@begingroup#1\color@endgroup}}
\patchcmd\@makecaption{\hbox}{\fix@ieeecolor@hbox}{}{\FAILED}
\patchcmd\@makecaption{\hbox}{\fix@ieeecolor@hbox}{}{\FAILED}

\usepackage{tmi}
\usepackage{cite}
\usepackage{amsmath,amssymb,amsfonts}
\usepackage{algorithmic}
\usepackage{graphicx}
\usepackage{textcomp}
\usepackage{amsmath}
\usepackage[switch]{lineno}

\def\BibTeX{{\rm B\kern-.05em{\sc i\kern-.025em b}\kern-.08em
T\kern-.1667em\lower.7ex\hbox{E}\kern-.125emX}}
\usepackage{algorithm}
\usepackage{array}
\usepackage{stfloats}
\usepackage{textcomp}
\usepackage{url}
\usepackage{graphicx}
\usepackage{verbatim}
\usepackage{cite}

\begin{document}

\title{Towards Modality- and Sampling-Universal Learning Strategies for Accelerating Cardiovascular Imaging: Summary of the CMRxRecon2024 Challenge}
\author{Fanwen Wang, Zi Wang, Yan Li, Jun Lyu, Chen Qin, Shuo Wang, Kunyuan Guo, Mengting Sun, Mingkai Huang, Haoyu Zhang, Michael Tänzer, Qirong Li, Xinran Chen, Jiahao Huang, Yinzhe Wu, Haosen Zhang, Kian Anvari Hamedani, Yuntong Lyu, Longyu Sun, Qing Li, Tianxing He, Lizhen Lan, Qiong Yao, Ziqiang Xu, Bingyu Xin, Dimitris N. Metaxas \IEEEmembership{Fellow, IEEE}, Narges Razizadeh, Shahabedin Nabavi, George Yiasemis, Jonas Teuwen, Zhenxi Zhang, Sha Wang, Chi Zhang, Daniel B. Ennis, Zhihao Xue, Chenxi Hu, Ruru Xu, Ilkay Oksuz, Donghang Lyu, Yanxin Huang, Xinrui Guo, Ruqian Hao, Jaykumar H. Patel, Guanke Cai, Binghua Chen, Sha Hua, Zhensen Chen, Qi Dou, Xiahai Zhuang \IEEEmembership{Senior Member, IEEE}, Qian Tao, Wenjia Bai \IEEEmembership{Senior Member, IEEE}, Jing Qin \IEEEmembership{Senior Member, IEEE}, He Wang, Claudia Prieto, Michael Markl, Alistair Young, Hao Li, Xihong Hu, Lianming Wu, Xiaobo Qu \IEEEmembership{Senior Member, IEEE}, Guang Yang \IEEEmembership{Senior Member, IEEE}, Chengyan Wang 
\thanks{In this paper, the co-first authors are Fanwen Wang, Zi Wang, Yan Li, Jun Lyu, and Chen Qin. The co-corresponding authors are Chengyan Wang, Xiaobo Qu, Guang Yang, Xihong Hu and Lianming Wu. Guang Yang and Chengyan Wang are co-last authors.}
\thanks{Fanwen Wang is with Shanghai Pudong Hospital and Human Phenome Institute, Fudan University, Shanghai, China; the Bioengineering Department and Imperial-X, Imperial College London, London, U.K and the Cardiovascular Magnetic Resonance Unit, Royal Brompton Hospital, London, U.K (e-mail: fanwen.wang@imperial.ac.uk).}
\thanks{Zi Wang, Xinran Chen and Haosen Zhang are with the Bioengineering Department and Imperial-X, Imperial College London, London, U.K (e-mail: zi.wang@imperial.ac.uk; sc2822@ic.ac.uk; haosen.zhang24@imperial.ac.uk).}
\thanks{Jiahao Huang and Yinzhe Wu are with the Bioengineering Department and Imperial-X, Imperial College London, London, U.K and the Cardiovascular Magnetic Resonance Unit, Royal Brompton Hospital, London, U.K. (e-mail: j.huang21@imperial.ac.uk; yinzhe.wu18@imperial.ac.uk).} 
\thanks{Yan Li is with the Department of Radiology, Ruijin Hospital, Shanghai Jiaotong University School of Medicine, Shanghai, China (e-mail: ly40730@rjh.com.cn).} 
\thanks{Jun Lyu is with Mass General Brigham, Harvard Medical School, MA, U.S.A. (e-mail: ljdream0710@pku.edu.cn).} 
\thanks{Chen Qin is with the Department of Electrical and Electronic Engineering \& I-X, Imperial College London, U.K. (e-mail: c.qin15@imperial.ac.uk).} 
\thanks{Shuo Wang is with the Digital Medical Research Center, School of Basic Medical Sciences, Fudan University, Shanghai, China (e-mail: shuowang@fudan.edu.cn).} 
\thanks{Kunyuan Guo and Mingkai Huang are with the Department of Electronic Science,Xiamen University-Neusoft Medical Magnetic Resonance Imaging Joint Research and Development Center, Fujian Provincial Key Laboratory of Plasma and Magnetic Resonance, Xiamen University, Xiamen, China (e-mail: keevinzha@stu.xmu.edu.cn; huangmingkai@stu.xmu.edu.cn).} 
\thanks{Mengting Sun, Longyu Sun, Qing Li, Tianxing He, Lizhen Lan and Haosen Zhang are with the Human Phenomics Institute, Fudan University, Shanghai, China (e-mail: sunmengting150@163.com; sunly22@m.fudan.edu.cn; 13188816530@163.com, tianxing\_he2001@163.com, lzlan20@fudan.edu.cn).} 
\thanks{Haoyu Zhang is with the Pen-Tung Sah Institute of Micro-Nano Science and Technology, Fujian Provincial Key Laboratory of Plasma and Magnetic Resonance, Xiamen University, Xiamen, China (e-mail: zhanghaoyu@stu.xmu.edu.cn).} 
\thanks{Michael Tänzer is with the Computing Department, Imperial College London, London, U.K and the Cardiovascular Magnetic Resonance Unit, Royal Brompton Hospital, London, U.K. (e-mail: m.tanzer@imperial.ac.uk).} 
\thanks{Qirong Li is with the School of Computer Science, Fudan University, Shanghai, China (e-mail: qrli21@m.fudan.edu.cn).} 
\thanks{Yuntong Lyu is with the School of Clinical Medicine, Zhongshan Hospital, Shanghai Medical College, Fudan University, Shanghai, China (e-mail: ytlv21@m.fudan.edu.cn).} 
\thanks{Ziqiang Xu is with Shanghai Fuying Medical Technology Co., Ltd., China (e-mail: daryl.xu@foxmail.com).} 
\thanks{Bingyu Xin and Dimitris N. Metaxas are with the Department of Computer Science, Rutgers University-New Brunswick, New Jersey, U.S.A. (e-mail: bx64@cs.rutgers.edu; dnm@cs.rutgers.edu).} 
\thanks{Kian Anvari Hamedani, Narges Razizadeh, and Shahabedin Nabavi are with the Faculty of Computer Science and Engineering, Shahid Beheshti University, Tehran, Iran(e-mail: k.anvarihamedani@mail.sbu.ac.ir; n.razizade@mail.sbu.ac.ir;s\_nabavi@sbu.ac.ir).} 
\thanks{George Yiasemis and Jonas Teuwen are with the Netherlands Cancer Institute, Amsterdam, Netherlands (e-mail: g.yiasemis@nki.nl; j.teuwen@nki.nl).}
\thanks{Zhenxi Zhang and Sha Wang are with Canon Medical Systems (China) Co., Ltd., Beijing, China (e-mail: zhenxi1.zhang@cn.medical.canon; sha.wang@cn.medical.canon).}
\thanks{Chi Zhang and Daniel B. Ennis are with the Department of Radiology, Stanford University, California, U.S.A. (e-mail: zcqlhf@stanford.edu; dbe@stanford.edu).}
\thanks{Zhihao Xue and Chenxi Hu are with the School of Biomedical Engineering, Shanghai Jiao Tong University, Shanghai, China (e-mail: thomasxue@sjtu.edu.cn; chenxi.hu@sjtu.edu.cn).}
\thanks{Ruru Xu and Ilkay Oksuz are with the Computer Engineering Department, Istanbul Technical University, Istanbul, Turkey (e-mail: xu21@itu.edu.tr; oksuzilkay@itu.edu.tr).}
\thanks{Donghang Lyu is with the Department of Radiology, Leiden University Medical Center, Leiden, Netherlands (e-mail: d.lyu@lumc.nl).}
\thanks{Yanxin Huang is with the School of Information and Communication Engineering, University of Electronic Science and Technology of China, Sichuan, China (e-mail: h1328121557@163.com).}
\thanks{Xinrui Guo is with the Wuhan National Laboratory for Optoelectronics, Huazhong University of Science and Technology, Wuhan, China (e-mail: xinruig@hust.edu.cn).}
\thanks{Ruqian Hao is with the School of Optoelectronic Science and Engineering, University of Electronic Science and Technology of China, Chengdu, China (e-mail: ruqian\_hao@uestc.edu.cn).}
\thanks{Jaykumar H. Patel is with the Department of Medical Biophysics and Physical Sciences, University of Toronto and Sunnybrook Research Institution, Toronto, Canada (e-mail: jaykumar.patel@mail.utoronto.ca).}
\thanks{Qiong Yao, Guanke Cai and Xihong Hu are with the Radiology Department, Children’s Hospital, Fudan University, Shanghai, China (e-mail: yaoqiong\_x\_ray@163.com; caiguanke2006@126.com; huxihong@fudan.edu.cn).}
\thanks{Binghua Chen and Lianming Wu are with the Department of Radiology, Ren Ji Hospital, School of Medicine, Shanghai Jiao Tong University, Shanghai, China (e-mail: chenbinghua0311@163.com; wulianming@shsmu.edu.cn).}
\thanks{Sha Hua is with the Department of Cardiovascular Medicine, Heart Failure Center, Ruijin Hospital Lu Wan Branch, Shanghai Jiao Tong University School of Medicine, Shanghai, China (e-mail: shahua@shsmu.edu.cn).}
\thanks{Zhensen Chen, He Wang and Hao Li are with the Institute of Science and Technology for Brain-Inspired Intelligence, Fudan University, Shanghai, China (e-mail: zhensenchen@fudan.edu.cn; hewang@fudan.edu.cn; h\_li@fudan.edu.cn).}
\thanks{Qi Dou is with the Department of Computer Science and Engineering, The Chinese University of Hong Kong, Hong Kong, China (e-mail: qidou@cuhk.edu.hk).}
\thanks{Xiahai Zhuang is with the School of Data Science, Fudan University, Shanghai, China (e-mail: zxh@fudan.edu.cn).}
\thanks{Qian Tao is with the Department of Imaging Physics, Delft University of Technology, the Netherlands (e-mail: q.tao@tudelft.nl).}
\thanks{Wenjia Bai is with the Department of Computing \& Department of Brain Sciences, Imperial College London, U.K. (e-mail: w.bai@imperial.ac.uk).}
\thanks{Jing Qin is with the School of Nursing, The Hong Kong Polytechnic University, Hong Kong, China (e-mail: harry.qin@polyu.edu.hk).}
\thanks{Claudia Prieto is with the School of Engineering and the iHEALTH Millenium Institute, Pontificia Universidad Católica de Chile, Santiago, Chile and School of Biomedical Engineering and Imaging Sciences, King’s College London, London, U.K. (e-mail: ccprieto@uc.cl).}
\thanks{Michael Markl is with the Department of Radiology, Feinberg School of Medicine, Northwestern University, Chicago, U.S.A. (e-mail: michael.markl@northwestern.edu).}
\thanks{Alistair Young is with the School of Biomedical Engineering and Imaging Sciences, King’s College London, U.K. (e-mail: alistair.young@kcl.ac.uk).}
\thanks{Xiaobo Qu is with the Department of Radiology, the First Affiliated Hospital of Xiamen University; the Department of Electronic Science, Xiamen University-Neusoft Medical Magnetic Resonance Imaging Joint Research and Development Center, Fujian Provincial Key Laboratory of Plasma and Magnetic Resonance, Xiamen University, Xiamen, China (e-mail: quxiaobo@xmu.edu.cn).} 
\thanks{Guang Yang is with Bioengineering Department, Imperial-X and National Heart and Lung Institute, Imperial College London, U.K.; Cardiovascular Magnetic Resonance Unit, Royal Brompton Hospital, London, U.K. and School of Biomedical Engineering and Imaging Sciences, King's College London, U.K. (email: g.yang@imperial.ac.uk).}
\thanks{Chengyan Wang is with Shanghai Pudong Hospital and Human Phenome Institute, Fudan University, Shanghai, China (e-mail: wangcy@fudan.edu.cn).}
\thanks{This study was supported in part by the Shanghai Municipal Science and Technology Major Project (no.2023SHZDZX02A05), the National Natural Science Foundation of China(no.62331021, 62371413, 62001120), the Shanghai Rising-Star Program (no.24QA2703300), the Scientific Research Fund Project of Pudong Hospital Affiliated to Fudan University (No.YJJC202409), the National Key R\&D Program of China (No.2024YFC3405800), the Specialty Feature Construction Project of Pudong Health and Family Planning Commission of Shanghai(No.PWZzb2022-29), the ERC IMI(101005122), the H2020 (952172), the MRC (MC/PC/21013), the Royal Society (IEC\textbackslash NSFC\textbackslash 211235), the NVIDIA Academic Hardware Grant Program, the SABER project supported by Boehringer Ingelheim Ltd, NIHR Imperial Biomedical Research Centre (RDA01), the Wellcome Leap Dynamic resilience program, UKRI guarantee funding for Horizon Europe MSCA Postdoctoral Fellowships(EP/Z002206/1), UKRI MRC Research Grant, TFS Research Grants (MR/U506710/1), Swiss National Science Foundation (Grant No. 220785), and the UKRI Future Leaders Fellowship (MR/V023799/1, UKRI2738), the Engineering and Physical Sciences Research Council UK Grants(no. EP/X039277/1, EP/Z533762/1), the Yantai Basic Research Key Project (no. 2023JCYJ041), the Youth Innovation Science and Technology Support Program of Shandong Provincial (no. 2023KJ239), and the Youth Program of Natural Science Foundation of Shandong Province (no. ZR2024QF001).}
}

\maketitle

\begin{abstract}
Cardiovascular health is vital to human well-being, and cardiac magnetic resonance (CMR) imaging is considered the {clinical reference standard} for diagnosing cardiovascular disease. However, its adoption is hindered by long scan times, complex contrasts, and inconsistent quality. While deep learning methods perform well on specific CMR imaging {sequences}, they often fail to generalize across modalities and sampling schemes. The lack of benchmarks for high-quality, fast CMR image reconstruction further limits technology comparison and adoption. The CMRxRecon2024 challenge, attracting over 200 teams from 18 countries, addressed these issues with two tasks: generalization to unseen {modalities} and robustness to diverse undersampling patterns. We introduced the largest public multi-{modality} CMR raw dataset, an open benchmarking platform, and shared code. Analysis of the best-performing solutions revealed that prompt-based adaptation and enhanced physics-driven consistency enabled strong cross-scenario performance. These findings establish principles for generalizable reconstruction models and advance clinically translatable AI in cardiovascular imaging.

\end{abstract}


\section{Introduction}
\label{sec:introduction}
Deep learning has driven rapid progress, yielding more accurate and efficient performance in medical tasks such as abnormality detection, image segmentation \cite{gatidis2024results}, image reconstruction \cite{dehner2023deep} and disease classification \cite{korot2021code}. Within healthcare, cardiovascular health stands out as a field benefiting a lot from such AI-driven innovations \cite{zhu2025sparse}. The total number of annual deaths from cardiovascular diseases (CVD) rose to 19.8 million in 2022, an increase from 12.4 million in 1990, and continues to account for approximately one-third of global mortality \cite{mensah2023global, global2023global}. Advanced and high-quality cardiac imaging is central to CVD diagnosis and management. While modalities like echocardiography and computed tomography provide valuable biomarkers, cardiovascular magnetic resonance (CMR) offers unique advantages as a non-invasive imaging tool with superior soft-tissue contrast and dynamic functional assessment capabilities \cite{la2012cardiac}. 

A standard CMR exam comprises multiple specialized imaging sequences with different contrasts, each tailored to evaluate specific aspects of cardiac structure, function, or tissue composition \cite{kramer2020standardized}. {In this work, we use the term modality to refer to individual CMR acquisition types, such as cine and mapping, that differ in their underlying pulse sequence and contrast mechanism.} The complementary CMR modalities provide a comprehensive evaluation of cardiac health, underscoring CMR’s critical role in both clinical cardiology and research \cite{petersen2016uk}. Despite its diagnostic value, multi-sequence CMR is time-consuming to acquire. Each CMR sequence often requires a separate breath-hold and precise cardiac gating, and collecting many slices or modalities in succession prolongs the overall scan time. Lengthy exams not only reduce patient comfort and throughput, but also increase the likelihood of motion artifacts that can degrade image quality. To address these challenges, accelerated imaging strategies encode the raw data (k-space) using various undersampling patterns and then employ specialized reconstruction algorithms to recover high-fidelity images from fewer measurements. Inspired by traditional model-based approaches, unrolled networks~\cite{sandino2021accelerating, kustner2020cinenet} integrate iterative optimization as a data consistency step while leveraging spatial and temporal convolutions for regularization. This architecture improves the efficiency of learning spatiotemporal priors, particularly in cine imaging. Additionally, deep priors capturing the smooth, low-dimensional manifold structure of cine images under radial undersampling have been explored~\cite{biswas2019dynamic}. Beyond unrolled networks, low-rank and sparse image models~\cite{huang2021deep,wang2024deepssl} and complementary time-frequency networks~\cite{qin2021complementary} were developed to tackle the challenges of dynamic cine acquisitions. In other CMR modalities, 3D U-Nets and complex-valued convolutional neural network (CNN) models were applied for 2D flow imaging reconstruction under real-time radial~\cite{haji2021highly}, spiral~\cite{jaubert2021deep}, and variable-density~\cite{cole2021analysis} sampling patterns. Complex convolutional networks were employed to reconstruct LGE images from Cartesian sampling~\cite{el2020deep} and recurrent CNNs with U-Net-refined maps demonstrated strong performance under Gaussian sampling patterns~\cite{jeelani2020myocardial}.

Despite this progress, most existing deep learning (DL)-based reconstruction models remain narrowly specialized to particular modalities, sampling patterns, or imaging protocols, which limits their utility across diverse clinical scenarios \cite{antun2020instabilities}. In practice, a full CMR study involves multiple different sequences, and deploying a separate tuned model for each modality imposes significant complexity, hindering real-time workflow integration \cite{wang2023pisf}. A key obstacle to a ``universal'' CMR reconstruction model is the substantial distribution shift between different sequence types and undersampling patterns: networks trained on one modality or k-space pattern often generalize poorly to others \cite{ouyang2019generalizing}. Furthermore, the training data available for each modality is typically limited in diversity. Existing public CMR datasets for reconstruction tend to either focus on a single modality with raw k-space data \cite{chen2020ocmr,elrewaidy2020multi}, or provide multi-modality collections of reconstructed images that lack corresponding raw multi-coil data \cite{zhuang2022cardiac,li2023myops,lalande2020emidec} (Table ~\ref{tab: datasets}). This scarcity of comprehensive multi-modality, multi-coil datasets hindered the development of models that can robustly handle a variety of CMR sequences and acquisition conditions.

\begin{table}[ht]
\centering
\caption{Overview of existing CMR datasets related to our dataset.}
\label{tab: datasets}
\scalebox{0.92}{
\begin{tabular}{|p{2.3cm}|p{0.4cm}|p{2.6cm}|p{1cm}|p{1cm}|}
\hline 
\textbf{Dataset Name} & \textbf{No. of Cases}& \textbf{Modality}& \textbf{Data Type} & \textbf{View}\\ \hline
\textbf{CMRxRecon2024~\cite{wang2024cmrxrecon2024}} & \textbf{330} & \textbf{Cine, Tagging, T1 mapping, T2 mapping, 2D-flow, Black-blood}& \textbf{k-space, Image} & \textbf{Multiple}\\ \hline
CMRxRecon2023~\cite{wang2024cmrxrecon}   & 300   & Cine, T1 mapping, T2 mapping & k-space, Image & Multiple\\ \hline
Harvard CMR Dataverse~\cite{elrewaidy2020multi} & 108 & Cine & k-space, Image & Multiple\\ \hline
OCMR~\cite{chen2020ocmr} & 53 & Cine & k-space, Image & Multiple\\ \hline
MS-CMR~\cite{zhuang2022cardiac}  & 45  & Cine, LGE, T2      & Image   & Multiple\\ \hline
MyoPS~\cite{li2023myops}   & 45   & Cine, LGE, T2      & Image & Single   \\ \hline
EMIDEC~\cite{lalande2020emidec} & 150  & Cine, Delay Enhanced & Image & Single   \\ \hline
M\&Ms~\cite{campello2021multi} & 375 & Cine & Image & Multiple\\ \hline
ACDC~\cite{bernard2018deep} & 150 & Cine & Image & Single \\ \hline
\end{tabular}
}
\end{table}

\begin{figure*}[t]
\centering
\includegraphics[width=\textwidth]{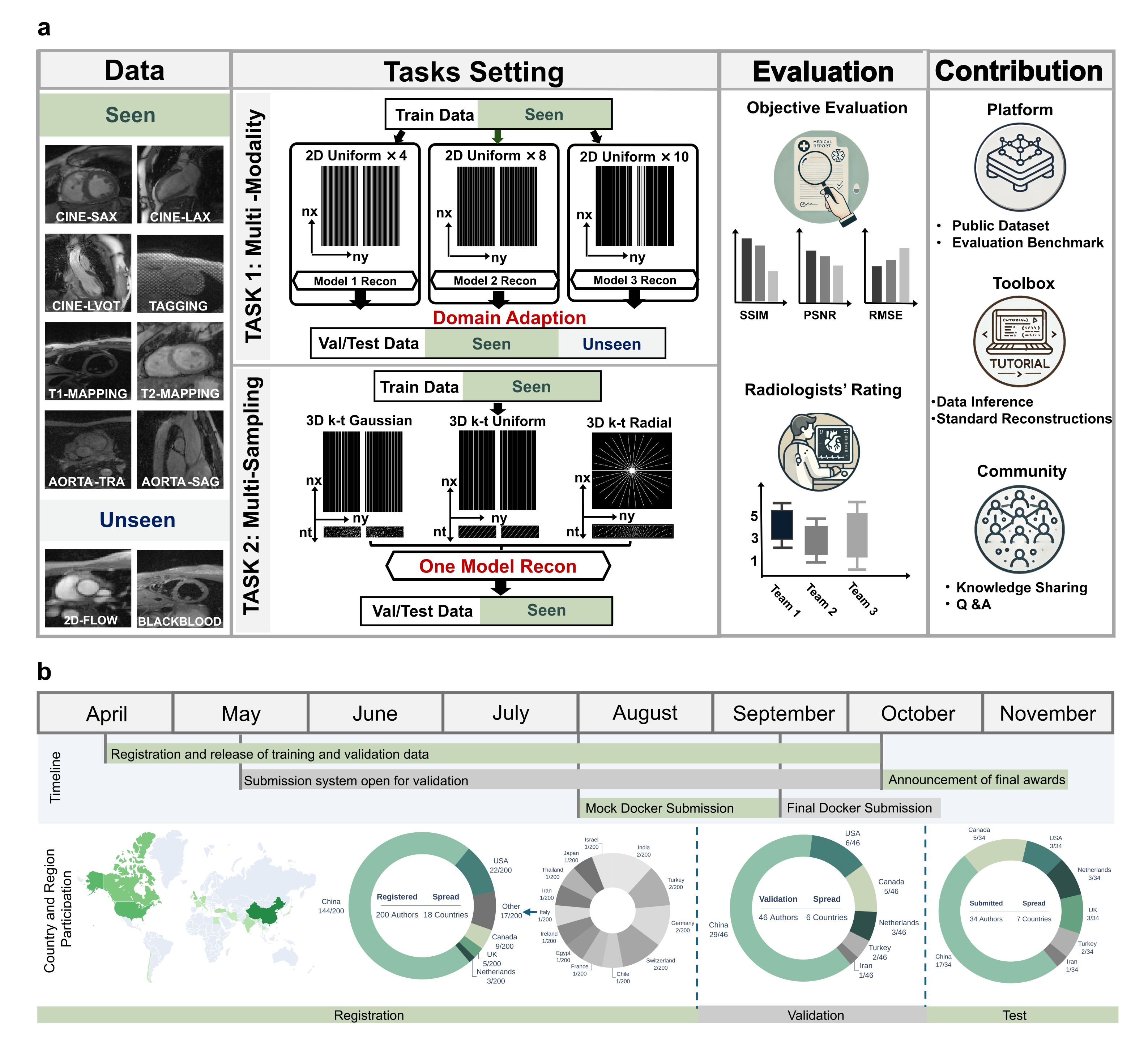}
\caption{\textbf{Challenge tasks, contributions and timeline.} \textbf{a.} This schematic illustrates the two challenge tasks with the data and modalities involved. Task 1 focuses on modality universality, training on seen modalities and testing on both seen (cine, tagging, T1 mapping, and T2 mapping) and unseen (black-blood and 2D flow) cases. Task 2 emphasizes sampling universality, and training on diverse 3D sampling schemes with a single model for inference. Evaluation includes objective metrics and radiologist ratings. Contributions include platform benchmarking, community evaluation criteria, and a toolbox with code resources. \textbf{b.} The timeline and summary of participation.}
\label{fig: challengeparticipation}
\end{figure*}

The CMRxRecon2024 challenge aimed to push forward the generalizable AI reconstruction for CMR by addressing these gaps (Fig. \ref{fig: challengeparticipation} a). Building on a previous challenge that provided 300 cases of cine and T1/T2 mapping data, CMRxRecon2024 expanded the scope to different CMR modalities (cine, 2D flow, T1 mapping, T2 mapping, tagging, and black-blood) across multiple views \cite{wang2024cmrxrecon,lyu2024state}. In total, it comprises over 200,000 k-space slices (roughly 2TB of raw data), all of which we have made publicly available via the Synapse platform. The challenge evaluates models on two fronts: reconstructing out-of-distribution CMR views/modalities, and maintaining robustness across unseen k-space sampling patterns and acceleration factors. This challenge framework aligns technical development with clinical needs: by improving scan efficiency and consistency across diverse conditions, such {modality- and sampling- universal} reconstruction approaches could streamline CMR workflows and make advanced imaging more accessible in practice.

In this work, we present the CMRxRecon2024 challenge, conducted as part of the Medical Image Computing and Computer-Assisted Intervention (MICCAI) 2024 conference, with the goals of: (1) advancing research on {modality- and sampling-universal} reconstruction models across diverse CMR modalities, views, and acceleration settings; (2) providing a publicly available multi-modality, multi-view CMR raw k-space dataset and standardized benchmarking platform for algorithm comparison and reproducibility, and (3) offering empirical insights into prompt-based model adaptation, dynamic coil sensitivity refinement, and the impact of model scale on generalization performance.

\section{Methods}
\subsection{Call for participation}
The mission of the CMRxRecon2024 challenge was to advance trustworthy, universal CMR image reconstruction – that is, to advance the development of reconstruction algorithms capable of handling a wide variety of imaging {modalities} and sampling schemes within a single model. To this end, the challenge defined two complementary tasks, each targeting a different aspect of generalizability (Fig. \ref{fig: challengeparticipation} a) :
(1) Task 1: Multi-modality CMR reconstruction. The goal of Task 1 was to develop a reconstruction model that could provide high-quality images for multiple CMR {modalities} types using one unified approach. Participants trained their models on k-space data from four common CMR modalities (cine, T1 mapping, T2 mapping, and tagging) and were required to reconstruct images from undersampled data of both these {four} seen modalities and two unseen modalities during testing (specifically, 2D phase-contrast flow and black-blood LGE, which were withheld from the training set). The challenge for Task 1 was thus to evaluate out-of-distribution generalization: a successful model would adapt to reconstructing new {modality} types with different image appearances and temporal dynamics that were not represented in its training data. Models in Task 1 focused on a modality-universal setting at fixed undersampling patterns – all input data used uniform Cartesian undersampling at acceleration rates of 4$\times$, 8$\times$, or 10$\times$ (each model was allowed to specialize to one acceleration factor). Participants could train separate models for each acceleration factor if desired, but each model had to handle all {modalities}. Performance was primarily judged on the model’s ability to reconstruct both the familiar and the unseen {modalities} with high fidelity.
(2) Task 2: Random sampling CMR reconstruction. Task 2 challenged participants to develop a single model robust to diverse k-space undersampling trajectories and acceleration factors. In this sampling-universal setting, the training data were the same four modalities as Task 1 with a variety of retrospectively applied undersampling patterns: Cartesian uniform masks, Cartesian Gaussian-variable density masks, and pseudo-radial masks, spanning acceleration factors from 4$\times$ up to 24$\times$. Participants were instructed to train one model that could handle all combinations of these trajectories and acceleration rates, reconstructing any of the three given {modalities} from any undersampled input. The final test cases for Task 2 covered the full range of sampling types and rates, including some that may have been especially scarce in training, to test generalization. This task evaluated a model’s ability to adapt to varying acquisition settings: e.g., maintaining performance when the undersampling pattern changes from one time frame to the next, or when acceleration is increased. Robust Task 2 models needed to implicitly recognize the sampling pattern and noise/artifact level in each input and adjust their reconstruction strategy accordingly to maintain image quality.
Each participating team could choose to tackle either or both tasks. 

\begin{table*}[t]
\centering
\caption{The list and details of the teams who successfully participated in the test (docker-submission) phase. M\# stands for the modality-universal Task 1. S\# stands for the sampling-universal Task 2.}
\begin{tabular}{|p{3cm}|p{10cm}|p{3cm}|} 
\hline
\textbf{Team} & \textbf{Affiliation} & \textbf{Location} \\ \hline 
M1/S1. CBIM~\cite{xin2024rethinking}   & Department of Computer Science, Rutgers University-New Brunswick               & New Jersey, U.S.A.      \\ \hline
M2. KNSynapse \cite{anvari2024all}  & Faculty of Computer Science and Engineering, Shahid Beheshti University            & Tehran, Iran       \\ \hline
M3/S2. direct \cite{yiasemis2022direct,yiasemis2024deep} & Netherlands Cancer Institute                                 & Amsterdam, Netherlands  \\ \hline
M4/S3. imr   & Canon Medical Systems (China) Co., Ltd.& Beijing, China      \\ \hline
S4. CardiAxs  & Department of Radiology, Stanford University                         & California, U.S.A.      \\ \hline
M5. SJTU\_CMR\_LAB & School of Biomedical Engineering, Shanghai Jiao Tong University               & Shanghai, China     \\ \hline
M6/S5. ITU PIMI Lab \cite{xu2024hypercmr} & Computer Engineering Department, Istanbul Technical University              & Istanbul, Turkey     \\ \hline
S6. LUMC~\cite{dlyu2025upcmr}    & Department of Radiology, Leiden University Medical Center                   & Leiden, Netherlands   \\ \hline
M7/S9. CMRxRecon2024-qiteam & School of Information and Communication Engineering, University of Electronic Science and Technology of China & Sichuan, China \\ \hline
S7. GuoLab   & Wuhan National Laboratory for Optoelectronics, Huazhong University of Science and Technology & Wuhan, China       \\ \hline
S8. MoemiCapy  & School of Optoelectronic Science and Engineering, University of Electronic Science and Technology of China & Sichuan, China \\ \hline
S10. SunnySD \cite{patel2024low} & Department of Medical Biophysics and Physical Sciences, University of Toronto and Sunnybrook Research Institution & Toronto, Canada \\ \hline
\end{tabular}
\label{tab: participant}
\end{table*}

\subsection{Data acquisition and preparation}
A new multi-{modality} CMR dataset, CMRxRecon2024, was collected and made publicly available for this challenge. Data were prospectively acquired with specifically designed multi-{modality} and multi-view protocols, using a 3T scanner (MAGNETOM Vida, Siemens Healthineers) equipped with dedicated multi-channel cardiac coils. Participants were positioned supine on the table before the scans. Electrodes were attached and electrocardiogram signals were recorded during the scanning process \cite{wang2021recommendation, wang2024cmrxrecon}. The dataset consists of raw k-space data from 330 healthy volunteer scans (156 males, 174 females, average aged 36 years) acquired under institutional review board approval (approval number: MS-R23). Our enrollment process and screening protocols can be found in \cite{wang2024cmrxrecon2024}. 

Data collection was conducted to cover six commonly used modalities with different anatomical views: (1) cine imaging with seven anatomical views, namely long-axis (LAX) (2-chamber, 3-chamber, and 4-chamber), short-axis (SAX), left ventricular outflow tract (LVOT), and aorta (transversal and sagittal views), (2) phase-contrast (i.e., 2D flow) with transversal view, (3) tagging with SAX view, (4) black-blood with SAX view, (5) T1 mapping with SAX view, and (6) T2 mapping with SAX view. {Specifically, in our data acquisition, a 2$\times$ parallel imaging acceleration was applied to shorten scan time, improve volunteer comfort, and mitigate motion artifacts. These low-acceleration undersampled k-space data were then reconstructed using GRAPPA \cite{griswold2002generalized} to generate the reference images. This setting is also commonly used in clinical CMR practice \cite{petersen2016uk,wang2024cmrxrecon}. These reference images were subsequently used as the benchmark for evaluating participating methods under different undersampling patterns and higher accelerations (4$\times$$\sim$24$\times$).} The detailed acquisition settings and parameters of imaging protocols are summarized in \cite{wang2024cmrxrecon2024}.

To generate different acceleration patterns in our challenge, various k-space undersampling trajectories (i.e., Cartesian uniform, Cartesian Gaussian, and pseudo radial) with different acceleration factors (AFs) (i.e., 4$\times$$\sim$24$\times$) were provided for retrospective k-space undersampling \cite{wang2024cmrxrecon2024}.

The collected multi-coil k-space data of these volunteers were split into three subsets: (1) training dataset with 200 individuals, (2) validation dataset with 60 individuals, and (3) test dataset with 70 individuals. The resulting CMRxRecon2024 dataset is openly accessible to individuals after challenge registration. 

\subsection{Challenge organization and infrastructure}
The challenge was organized as an official MICCAI 2024 competition and was conducted from April to {October} 2024. Training and validation data were released in April 2024, giving participants several months to develop and tune their models. A validation leaderboard was available to teams to get feedback on their performance. {During the validation phase, participants receive detailed feedback automatically generated by the evaluation script. The feedback includes, for each case and modality: (1) the quantitative metrics (PSNR, SSIM, and NMSE), (2) the success rate summarizing the proportion of valid reconstructions, and (3) diagnostic messages indicating the reason for any failure, such as missing files, dimension mismatches, or invalid numerical values (NaN/Inf).} In September 2024, participants submitted their final containerized algorithms which were run on a secure server to reconstruct the private test set. The entire evaluation was managed on the Synapse platform, which provided a standardized environment and ensured fair comparison. To foster collaboration and transparency, the organizers provided a code library with utilities for k-space manipulation and data loading. {Participants were allowed to train their models on our provided CMRxRecon2024 dataset \cite{wang2024cmrxrecon2024} from scratch, or to use the model pretrained on two public datasets (fastMRI \cite{knoll2020fastmri} and CMRxRecon2023 \cite{wang2024cmrxrecon}. Any private or unpublished datasets were strictly prohibited. To ensure auditability and compliance, each team was required to provide a detailed report of their training procedure, including whether pretraining was used and the corresponding data sources. We manually checked all submitted reports to verify adherence to the rules. Since all permitted datasets are public, their use is transparent and verifiable, and any suspicious or undeclared use of non-compliant data would have been flagged and disqualified. These measures collectively ensured fairness among all participating teams.} Teams were also encouraged to publish their methods in short workshop papers. After the final phase, an in-person workshop session was held at MICCAI 2024 where the challenge results were presented, and the winning teams discussed their approaches. Prizes were awarded to the top five teams of each task. 

All docker submissions were executed on the same Linux workstation equipped with an Intel(R) Xeon(R) E5-2698 v4 processor (2.20GHz base frequency, 40 cores), 256GB of memory, and one NVIDIA Tesla V100-DGXS-32GB graphics processor. 

\begin{table}[t]
\caption{Ranking Table for Task 1 (p-value: * $<$ 0.05, ** $<$ 0.01). "NA" in the row of reference indicates they work as the reference, while "NA" in the column of the Rad. Score means the teams beyond the top five are not ranked. {The value in the parenthesis represents the standard deviation.}}
\scalebox{0.69}{
\begin{tabular}{|l|l|l|l|l|l|}
\hline
\textbf{Team} & \multicolumn{1}{l|}{\textbf{SSIM}} & \multicolumn{1}{l|}{\textbf{PSNR}} & \multicolumn{1}{l|}{\textbf{NMSE}} & \multicolumn{1}{l|}{\textbf{Rad. Score}} & \multicolumn{1}{l|} {\textbf{Rank}}\\
\hline
\includegraphics[width=0.015\textwidth]{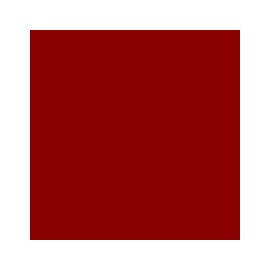} \text{M1. CBIM} & 0.980 (0.009) & 44.80 (2.50) & 0.007 (0.004) & 4.82 (0.11) & 1 \\ \hline
\includegraphics[width=0.015\textwidth]{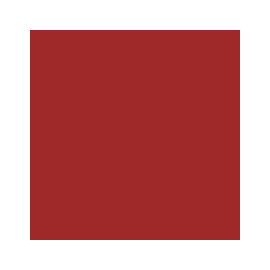} \text{M2. KNSynapse} & 0.978 (0.010)$^{**}$ & 44.02 (2.83)$^{**}$ & 0.008 (0.005)$^{**}$ & 4.80 (0.13) & 2\\ \hline
\includegraphics[width=0.015\textwidth]{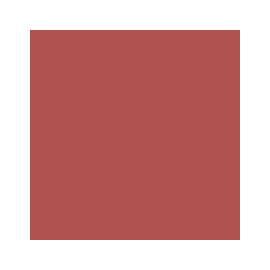} \text{M3. direct} & 0.977 (0.009)$^{**}$ & 43.94 (2.50)$^{**}$ & 0.008 (0.004)$^{**}$ & 4.80 (0.13) & 3 \\ \hline
\includegraphics[width=0.015\textwidth]{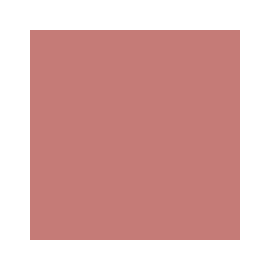} \text{M4. imr} & 0.977 (0.011)$^{**}$ & 43.71 (3.09)$^{**}$ & 0.009 (0.005)$^{*}$ & 4.79 (0.11) & 3 \\ \hline
\includegraphics[width=0.015\textwidth]{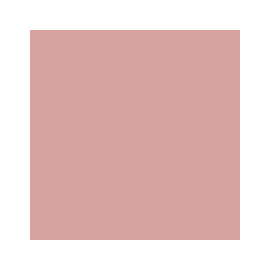} \text{M5. SJT\_CMR\_LAB} & 0.964 (0.015)$^{**}$ & 41.30 (3.16)$^{**}$ & 0.015 (0.008)$^{**}$ & 4.74 (0.12) & 5 \\ \hline
\includegraphics[width=0.015\textwidth]{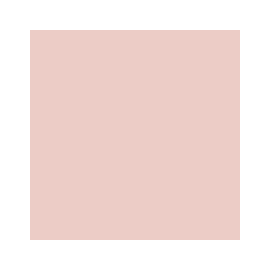} \text{M6. ITU PIMI Lab} & 0.964 (0.017)$^{**}$ & 41.76 (3.19)$^{**}$ & 0.013 (0.008)$^{**}$ & \text{NA} (\text{NA}) & 6 \\ \hline
\includegraphics[width=0.015\textwidth]{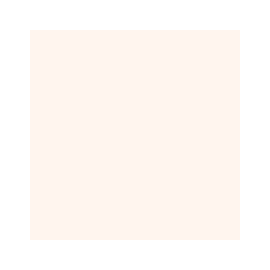} \text{M7. qi-team} & 0.963 (0.018)$^{**}$ & 41.19 (3.46)$^{**}$ & 0.016 (0.010)$^{**}$ & \text{NA} (\text{NA}) & 7 \\ \hline
\text{Reference} & \text{NA} (\text{NA}) & \text{NA} (\text{NA}) & \text{NA} (\text{NA}) & 4.53 (0.19) & \text{NA}\\ \hline
\end{tabular}}
\label{tab: task1rank}
\end{table}

\begin{table}[t]
\caption{Ranking Table for Task 2 (p-value: * $<$ 0.05, ** $<$ 0.01). "NA" in row of reference indicates they work as the reference, while "NA" in the column of the Rad. Score means the teams beyond the top five are not ranked. {The value in the parenthesis represents the standard deviation.}}
\scalebox{0.7}{
\begin{tabular}{|l|l|l|l|l|l|}
\hline
\textbf{Team} & \multicolumn{1}{l|}{\textbf{SSIM}} & \multicolumn{1}{l|}{\textbf{PSNR}} & \multicolumn{1}{l|}{\textbf{NMSE}} & \multicolumn{1}{l|}{\textbf{Rad. Score}} & \multicolumn{1}{l|} {\textbf{Rank}}\\
\hline
\includegraphics[width=0.015\textwidth]{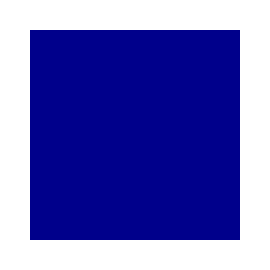} \text{S1. CBIM} & 0.977 (0.005) & 43.33 (1.22) & 0.009 (0.002) & 4.86 (0.13) & 1 \\ \hline
\includegraphics[width=0.015\textwidth]{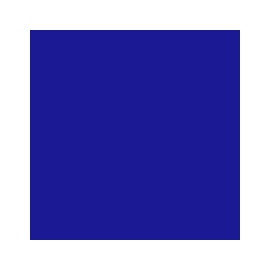} \text{S2. direct} & 0.974 (0.007)$^{*}$ & 42.58 (1.65)$^{**}$ & 0.011 (0.004)$^{**}$ & 4.83 (0.12) & 2 \\ \hline
\includegraphics[width=0.015\textwidth]{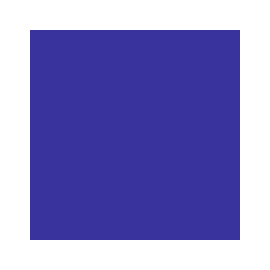} \text{S3. imr} & 0.970 (0.007)$^{**}$ & 41.69 (1.44)$^{**}$ & 0.013 (0.004)$^{**}$ & 4.86 (0.12) & 2 \\ \hline
\includegraphics[width=0.015\textwidth]{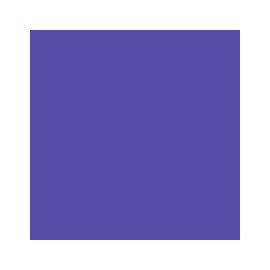} \text{S4. CardiAxs} & 0.954 (0.011)$^{**}$ & 39.47 (1.70)$^{**}$ & 0.019 (0.007)$^{**}$ & 4.74 (0.20) & 4\\ \hline
\includegraphics[width=0.015\textwidth]{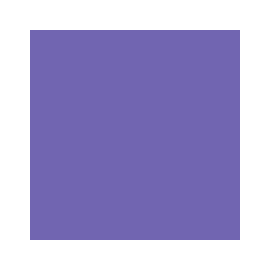} \text{S5. ITU PIMI Lab} & 0.947 (0.016)$^{**}$ & 38.76 (2.21)$^{**}$ & 0.025 (0.010)$^{**}$ & 4.72 (0.20) & 5\\ \hline
\includegraphics[width=0.015\textwidth]{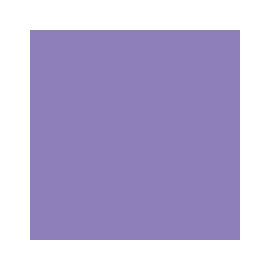} \text{S6. LUMC} & 0.921 (0.012)$^{**}$ & 33.57 (0.71)$^{**}$ & 0.065 (0.007)$^{**}$ & \text{NA} (\text{NA}) & 6\\ \hline
\includegraphics[width=0.015\textwidth]{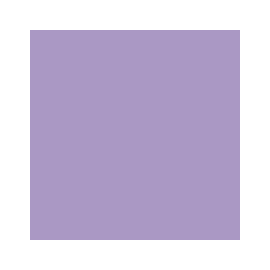} \text{S7. GUO\_LAB} & 0.903 (0.029)$^{**}$ & 34.23 (2.15)$^{**}$ & 0.059 (0.022)$^{**}$ & \text{NA} (\text{NA}) & 7\\ \hline
\includegraphics[width=0.015\textwidth]{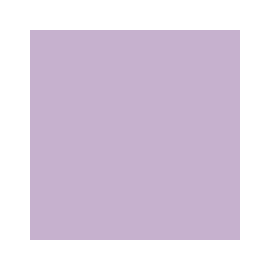} \text{S8. MoemilCapy} & 0.773 (0.058)$^{**}$ & 29.19 (1.60)$^{**}$ & 0.135 (0.032)$^{**}$ & \text{NA} (\text{NA}) & 8\\ \hline
\includegraphics[width=0.015\textwidth]{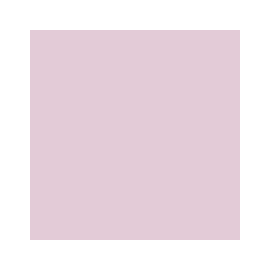} \text{S9. qi-team} & 0.750 (0.058)$^{**}$ & 28.60 (1.71)$^{**}$ & 0.208 (0.056)$^{**}$ & \text{NA} (\text{NA}) & 9\\ \hline
\includegraphics[width=0.015\textwidth]{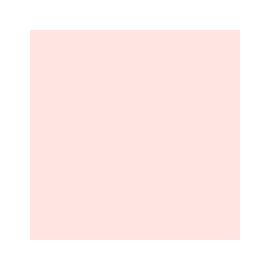} \text{S10. SunnySD} & 0.687 (0.013)$^{**}$ & 30.60 (0.64)$^{**}$ & 0.306 (0.028)$^{**}$ & \text{NA} (\text{NA}) & 10 \\ \hline
\text{Reference} & \text{NA} (\text{NA}) & \text{NA} (\text{NA}) & \text{NA} (\text{NA}) & 4.47 (0.26) & \text{NA}\\
\hline
\end{tabular}}
\label{tab: task2rank}
\end{table}

\subsection{Challenge evaluation metrics}
We {employed} two evaluation criteria: objective metrics and radiologists' rankings. The objective quantitative metrics include Structural Similarity Index (SSIM), Peak Signal-to-Noise Ratio (PSNR), and Normalized Mean Squared Error (NMSE). {For each modality and undersampling pattern, quantitative metrics were first computed on every reconstructed volume whose output dimensionality exactly matched that of the corresponding reference. A reconstruction was considered successful only if its array shape was identical to the reference and the data contained finite numerical values (no NaN/Inf). Outputs failing these checks were labeled as invalid and excluded from the metric computation of successful cases.}

Let $n$ denote the number of successfully reconstructed cases and $N$ the total number of available cases within a modality. The success rate was then defined as:
\begin{equation}
w = \frac{n}{N},
\label{eq:weight}
\end{equation}
{To prevent models that reconstructed only a few easy samples from obtaining disproportionately high average metrics, each modality-specific mean metric was further weighted by its success rate.}

Specifically, for PSNR and SSIM, the adjusted values were computed as
\begin{equation}
\mathrm{PSNR}_{\mathrm{adj}} = w \times \mathrm{PSNR}_{\mathrm{mean}}, \quad
\mathrm{SSIM}_{\mathrm{adj}} = w \times \mathrm{SSIM}_{\mathrm{mean}},
\end{equation}
and for NMSE, which is a lower-is-better metric,
\begin{equation}
\mathrm{NMSE}_{\mathrm{adj}} = (2 - w) \times \mathrm{NMSE}_{\mathrm{mean}},
\end{equation}
{where $w \in [0,1]$ denotes the success rate. This formulation ensures that the adjusted NMSE remains within a bounded penalty range} $[\,\mathrm{NMSE}_{\mathrm{mean}},\,2\times\mathrm{NMSE}_{\mathrm{mean}}\,]$, doubling the value in the worst case ($w=0$) and preserving the original score when all reconstructions are valid ($w=1$).

{Because $w$ is dimensionless, multiplying it by a statistical metric remains dimensionally valid. The weighting was applied within each modality before averaging across modalities and sampling patterns, ensuring equal contribution of all categories in the final ranking.}

Three independent radiologists {rated the central slice of each modality, sampling pattern, and reconstruction for ten random subjects per task, resulting in $>$ 1,000 individual image evaluations per reader (Task 1: 1,230; Task 2: 1,105). Each Task 1 subject contained 3 sampling patterns (Uniform $\times$4, $\times$8, $\times$10) $\times$ 10 modalities and views $\times$ 6 reconstruction sources (5 teams + reference); each Task 2 case contained 3 sampling patterns (Gaussian, Uniform, Radial; AF = 4$\times$$\sim$24$\times$) $\times$ 8 modalities and views $\times$ 6 sources. }

Image quality was rated on a five-point scale: 5 (excellent), 4 (good), 3 (fair), 2 (poor), and 1 (non-diagnostic). Evaluations considered artifacts, signal-to-noise ratio, and texture inconsistencies. Scores from each radiologist are first averaged for each sampling pattern and acceleration factor. Subsequently, the results were averaged across the three radiologists. {To balance the radiologists’ subjective scores and the objective reconstruction metrics, the final ranking for both tasks was determined by averaging the rankings derived from SSIM and the radiologists’ scores.}


For statistical comparison, paired two-sided Wilcoxon signed-rank tests were conducted to assess whether the performance differences between each team and the top-ranked team were significant. The tests were performed per case and per metric, and the resulting p-values indicate the significance of these paired differences (p-value: * $<$ 0.05, ** $<$ 0.01).

Following \cite{lyu2024state}, we evaluated the computational efficiency based on {RAM, GPU RAM, model parameters, runtime, per volume and per frame latencies with I/O overheads, throughput (slices/s) and energy estimates}.

\begin{figure}[t]
\centering
\includegraphics[width=\columnwidth]{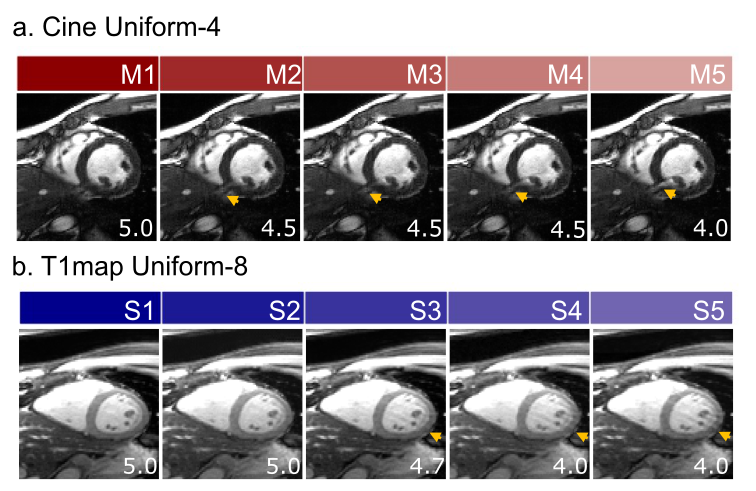}
\caption{{\textbf{CMRxRecon2024 Challenge Results.} Representative reconstructed images follow the ranking results of the top five teams in Task 1 and Task 2 respectively. The mean radiologists' rating for each reconstruction is displayed in the bottom-right corner of each image. Yellow arrows highlight undesired artifacts. }}
\label{fig: challenge_figure}
\end{figure}

\subsection{Data availability}
We provided the largest multi-modality and multi-view CMR raw k-space dataset, comprising 0.2 million sections of k-space data totaling over 2 TB available on Synapse (\textit{https://www.synapse.org/Synapse:syn54951257/wiki/627141}). 
The dataset is accessible {after registration} at: \textit{https://www.synapse.org/Synapse:syn54951257/wiki/627141}, {with license as CC BY-NC-ND}. The leaderboard and detailed performance metrics can be viewed at: \textit{https://www.synapse.org/Synapse:syn54951257/wiki/627936}. The challenge proceedings can be accessed through: \textit{https://link.springer.com/book/10.1007/978-3-031-87756-8}

\subsection{Code availability}
All codes, tutorials, data processing, and model pool are publicly available via GitHub, under an MIT license, at: \textit{https://github.com/CmrxRecon/CMRxRecon2024}.

\section{Results}
In the following section, we describe the challenge setting and participation, overall ranking and highlights of the two challenge tasks.

\subsection{Challenge setting and participation}

The challenge (\textit{https://cmrxrecon.github.io/2024/Home.html}) spanned from April to {October} 2024 (Fig. \ref{fig: challengeparticipation} b). The CMRxRecon2024 challenge saw a world-wide level of participation. In total, 200 teams from 18 different countries registered for the challenge. After an initial validation phase, 46 teams proceeded to the final testing phase by successfully submitting dockerized reconstruction algorithms on the evaluation platform. Ultimately, {12} teams (spanning North America, Europe, and Asia) completed the final phase with valid results on both tasks. The reconstruction results of these teams were evaluated on a hidden test set and ranked according to objective image quality metrics. The details of all participating teams are summarized in the Table \ref{tab: participant}.

\begin{table*}[t]
\centering
\caption{{Characteristics of models of all ranked teams in Task 1. Abbreviations: flip (F), rotation (R), shift (S), data consistency (DC), gradient descent (GD), learning rate (LR), not applicable (NA).}}
\scalebox{0.65}{
\begin{tabular}{|l|lllll|lllll|lll|}
\hline
\textbf{Team} & \multicolumn{5}{l|}{\textbf{Data processing}} & \multicolumn{5}{l|}{\textbf{Model information}} & \multicolumn{3}{l|}{\textbf{Training configuration}} \\ \cline{2-14} 
& \multicolumn{1}{l|}{Standardization} & \multicolumn{4}{l|}{Augmentation} & \multicolumn{1}{l|}{Network backbone} & \multicolumn{2}{l|}{Physical model} & Modality fusion & \multicolumn{1}{|l|}{Unrolled} & \multicolumn{1}{l|}{Optimizer / LR} & \multicolumn{1}{l|}{GPU hardware} & \multicolumn{1}{l|}{External pretraining dataset} \\ \cline{3-6} \cline{8-9} 
& \multicolumn{1}{l|}{} & \multicolumn{1}{l|}{F} & \multicolumn{1}{l|}{R} & \multicolumn{1}{l|}{S} & Others & \multicolumn{1}{l|}{} & \multicolumn{1}{l|}{DC} & \multicolumn{1}{l|}{Others} & \multicolumn{1}{l|}{} & \multicolumn{1}{l|}{} & \multicolumn{1}{l|}{} & \multicolumn{1}{l|}{} & \multicolumn{1}{l|}{} \\ \hline
M1. CBIM & \multicolumn{1}{l|}{Z-score} & \multicolumn{1}{l|}{} & \multicolumn{1}{l|}{} & \multicolumn{1}{l|}{} & N/A & \multicolumn{1}{l|}{Prompt-UNet~\cite{potlapalli2023promptir}} & \multicolumn{1}{l|}{\checkmark} & \multicolumn{1}{l|}{GD} & Channel attention & \multicolumn{1}{|l|}{\checkmark} & \multicolumn{1}{l|}{AdamW / 2e-4} & \multicolumn{1}{l|}{4$\times$A100 (80 GB)} & None \\ \hline
M2. KNSynapse & \multicolumn{1}{l|}{Z-score} & \multicolumn{1}{l|}{} & \multicolumn{1}{l|}{} & \multicolumn{1}{l|}{} & Data balancing & \multicolumn{1}{l|}{Prompt-UNet~\cite{potlapalli2023promptir}} & \multicolumn{1}{l|}{\checkmark} & \multicolumn{1}{l|}{GD} & Modality prompt & \multicolumn{1}{|l|}{\checkmark} & \multicolumn{1}{l|}{AdamW / 2e-3} & \multicolumn{1}{l|}{2$\times$H100 (80 GB)} & 80 subjects from CMRxRecon2023~\cite{wang2024cmrxrecon} \\ \hline
M3. direct & \multicolumn{1}{l|}{Max} & \multicolumn{1}{l|}{\checkmark} & \multicolumn{1}{l|}{} & \multicolumn{1}{l|}{\checkmark} & N/A & \multicolumn{1}{l|}{vSHARP~\cite{yiasemis2025vsharp}} & \multicolumn{1}{|l|}{\checkmark} & \multicolumn{1}{l|}{ADMM} & Network refining & \multicolumn{1}{|l|}{\checkmark} & \multicolumn{1}{l|}{Adam / 1.6e-4} & \multicolumn{1}{l|}{H100 (80 GB)} & None \\ \hline
M4. imr & \multicolumn{1}{l|}{Z-score} & \multicolumn{1}{l|}{} & \multicolumn{1}{l|}{} & \multicolumn{1}{l|}{} & N/A & \multicolumn{1}{l|}{E2E-VarNet~\cite{sriram2020end}} & \multicolumn{1}{l|}{\checkmark} & \multicolumn{1}{l|}{GD} & N/A & \multicolumn{1}{|l|}{\checkmark} & \multicolumn{1}{l|}{AdamW / 1e-4} & \multicolumn{1}{l|}{A800 (80 GB)} & None \\ \hline
M5. SJTU-CMR & \multicolumn{1}{l|}{Max} & \multicolumn{1}{l|}{} & \multicolumn{1}{l|}{} & \multicolumn{1}{l|}{} & N/A & \multicolumn{1}{l|}{E2E-VarNet~\cite{sriram2020end}} & \multicolumn{1}{l|}{\checkmark} & \multicolumn{1}{l|}{GD} & N/A & \multicolumn{1}{|l|}{\checkmark} & \multicolumn{1}{l|}{Adam / 5e-4} & \multicolumn{1}{l|}{RTX3090 (24 GB)} & None \\ \hline
M6. ITU-PIMI & \multicolumn{1}{l|}{Min-Max} & \multicolumn{1}{l|}{\checkmark} & \multicolumn{1}{l|}{} & \multicolumn{1}{l|}{\checkmark} & Image cropping & \multicolumn{1}{l|}{Prompt-UNet~\cite{potlapalli2023promptir}} & \multicolumn{1}{l|}{\checkmark} & \multicolumn{1}{l|}{GD} & Layer sharing & \multicolumn{1}{|l|}{\checkmark} & \multicolumn{1}{l|}{AdamW / 1e-4} & \multicolumn{1}{l|}{RTX3090 (24 GB)} & None \\ \hline
M7. qiteam & \multicolumn{1}{l|}{Min-Max} & \multicolumn{1}{l|}{\checkmark} & \multicolumn{1}{l|}{\checkmark} & \multicolumn{1}{l|}{\checkmark} & Image cropping & \multicolumn{1}{l|}{Prompt-UNet~\cite{potlapalli2023promptir}} & \multicolumn{1}{l|}{\checkmark} & \multicolumn{1}{l|}{ADMM} & Feature fusion & \multicolumn{1}{|l|}{} & \multicolumn{1}{l|}{Adam / 2e-4} & \multicolumn{1}{l|}{A100 (40 GB)} & None \\ \hline
\end{tabular}
}
\label{tab: bullet_task1}
\end{table*}

\begin{table*}[]
\centering
\caption{{Characteristics of models for all ranked teams in Task 2. Abbreviations: flip (F), rotation (R), shift (S), data consistency (DC), gradient descent (GD), conjugate gradient (CG), learning rate (LR), not applicable (NA).}}
\scalebox{0.65}{
\begin{tabular}{|l|lllll|lllll|lll|}
\hline
\textbf{Team} & \multicolumn{5}{l|}{\textbf{Data processing}} & \multicolumn{5}{l|}{\textbf{Model information}} & \multicolumn{3}{l|}{\textbf{Training configuration}} \\ \cline{2-14} 
& \multicolumn{1}{l|}{Standardization} & \multicolumn{4}{l|}{Augmentation} & \multicolumn{1}{l|}{Network backbone} & \multicolumn{2}{l|}{Physical model} & Modality fusion & \multicolumn{1}{|l|}{Unrolled} & \multicolumn{1}{l|}{Optimizer / LR} & \multicolumn{1}{l|}{GPU hardware} & \multicolumn{1}{l|}{External pretraining dataset} \\ \cline{3-6} \cline{8-9} 
& \multicolumn{1}{l|}{} & \multicolumn{1}{l|}{F} & \multicolumn{1}{l|}{R} & \multicolumn{1}{l|}{S} & Others & \multicolumn{1}{l|}{} & \multicolumn{1}{l|}{DC} & \multicolumn{1}{l|}{Others} & \multicolumn{1}{l|}{} & \multicolumn{1}{l|}{} & \multicolumn{1}{l|}{} & \multicolumn{1}{l|}{} & \multicolumn{1}{l|}{} \\ \hline
S1. CBIM    & \multicolumn{1}{l|}{Z-score} & \multicolumn{1}{l|}{}      & \multicolumn{1}{l|}{}      & \multicolumn{1}{l|}{}      & N/A       & \multicolumn{1}{l|}{Prompt-UNet~\cite{potlapalli2023promptir}} & \multicolumn{1}{l|}{\checkmark} & \multicolumn{1}{l|}{GD} & Channel attention & \multicolumn{1}{|l|}{\checkmark} & \multicolumn{1}{l|}{AdamW / 2e-4} & \multicolumn{1}{l|}{4$\times$A100 (80 GB)} & None \\ \hline
S2. direct   & \multicolumn{1}{l|}{Max}   & \multicolumn{1}{l|}{\checkmark} & \multicolumn{1}{l|}{}      & \multicolumn{1}{l|}{\checkmark} & N/A       & \multicolumn{1}{l|}{vSHARP \cite{yiasemis2025vsharp}} & \multicolumn{1}{l|}{\checkmark} & \multicolumn{1}{l|}{ADMM} & Network refining & \multicolumn{1}{|l|}{\checkmark} & \multicolumn{1}{l|}{Adam / 1.6e-4} & \multicolumn{1}{l|}{H100 (80 GB)} & None \\ \hline
S3. imr    & \multicolumn{1}{l|}{Z-score} & \multicolumn{1}{l|}{}      & \multicolumn{1}{l|}{}      & \multicolumn{1}{l|}{}      & N/A       & \multicolumn{1}{l|}{E2E-VarNet \cite{sriram2020end}} & \multicolumn{1}{|l|}{\checkmark} & \multicolumn{1}{l|}{GD} & N/A    & \multicolumn{1}{|l|}{\checkmark}   & \multicolumn{1}{l|}{AdamW / 1e-4} & \multicolumn{1}{l|}{A800 (80 GB)} & None \\ \hline
S4. CardiAxs  & \multicolumn{1}{l|}{Max}   & \multicolumn{1}{l|}{}      & \multicolumn{1}{l|}{}      & \multicolumn{1}{l|}{}      & N/A       & \multicolumn{1}{l|}{Prompt-UNet~\cite{potlapalli2023promptir}} & \multicolumn{1}{l|}{\checkmark} & \multicolumn{1}{l|}{GD} & Modality prompt & \multicolumn{1}{|l|}{\checkmark} & \multicolumn{1}{l|}{Adam / 5e-5} & \multicolumn{1}{l|}{P40 (24 GB)} & 80 subjects from CMRxRecon2023~\cite{wang2024cmrxrecon} \\ \hline
S5. ITU-PIMI  & \multicolumn{1}{l|}{Min-Max} & \multicolumn{1}{l|}{\checkmark} & \multicolumn{1}{l|}{}      & \multicolumn{1}{l|}{\checkmark} & Image cropping  & \multicolumn{1}{l|}{Prompt-UNet~\cite{potlapalli2023promptir}} & \multicolumn{1}{l|}{\checkmark} & \multicolumn{1}{l|}{GD} & Layer sharing  & \multicolumn{1}{|l|}{\checkmark} & \multicolumn{1}{|l|}{AdamW / 3e-4} & \multicolumn{1}{l|}{4$\times$A100 (80 GB)} & None \\ \hline
S6. LUMC    & \multicolumn{1}{l|}{Max}   & \multicolumn{1}{l|}{}      & \multicolumn{1}{l|}{}      & \multicolumn{1}{l|}{}      & N/A       & \multicolumn{1}{l|}{UNet \cite{falk2019u}} & \multicolumn{1}{|l|}{\checkmark} & \multicolumn{1}{l|}{ADMM} & N/A & \multicolumn{1}{|l|}{\checkmark}       & \multicolumn{1}{l|}{AdamW / 2e-4} & \multicolumn{1}{l|}{A100 (80 GB)} & None \\ \hline
S7. GuoLab   & \multicolumn{1}{l|}{Min-Max} & \multicolumn{1}{l|}{}      & \multicolumn{1}{l|}{}      & \multicolumn{1}{l|}{}      & N/A       & \multicolumn{1}{l|}{ResNet \cite{He_2016_CVPR}} & \multicolumn{1}{l|}{\checkmark} & \multicolumn{1}{l|}{GD} & N/A    & \multicolumn{1}{|l|}{\checkmark}    & \multicolumn{1}{l|}{Adam / 1e-3} & \multicolumn{1}{l|}{RTX4090 (24 GB)} & None \\ \hline
S8. MoemilCapy & \multicolumn{1}{l|}{Min-Max} & \multicolumn{1}{l|}{}      & \multicolumn{1}{l|}{}      & \multicolumn{1}{l|}{}      & k-space padding & \multicolumn{1}{l|}{GNA-UNet \cite{dietlmeier2023cardiac}} & \multicolumn{1}{l|}{}     & \multicolumn{1}{l|}{N/A} & N/A    & \multicolumn{1}{|l|}{\checkmark}   & \multicolumn{1}{l|}{AdamW / 1e-4} & \multicolumn{1}{l|}{RTX3090 (24 GB)} & None \\ \hline
S9. qiteam   & \multicolumn{1}{l|}{Min-Max} & \multicolumn{1}{l|}{\checkmark} & \multicolumn{1}{l|}{\checkmark} & \multicolumn{1}{l|}{\checkmark} & Image cropping  & \multicolumn{1}{l|}{Prompt-UNet~\cite{potlapalli2023promptir}} & \multicolumn{1}{l|}{\checkmark} & \multicolumn{1}{l|}{ADMM} & Feature fusion & \multicolumn{1}{|l|}{} & \multicolumn{1}{l|}{Adam / 2e-4} & \multicolumn{1}{l|}{A100 (40 GB)} & None \\ \hline
S10. SunnySD  & \multicolumn{1}{l|}{Max}   & \multicolumn{1}{l|}{}      & \multicolumn{1}{l|}{}      & \multicolumn{1}{l|}{}      & Random matching & \multicolumn{1}{l|}{UNet \cite{falk2019u}} & \multicolumn{1}{l|}{\checkmark} & \multicolumn{1}{l|}{CG} & Low-rank basis & \multicolumn{1}{|l|}{} & \multicolumn{1}{l|}{Adam / 1e-3} & \multicolumn{1}{l|}{4$\times$P100 (16 GB)} & None \\ \hline
\end{tabular}
}
\label{tab: bullet_task2}
\end{table*}

\subsection{Overall ranking of the challenge}

The rankings for Task 1 and Task 2 are shown in Table. \ref{tab: task1rank} and Table. \ref{tab: task2rank} respectively. {The objective SSIM metrics showed a strong alignment with the radiologists’ ratings, with only minor deviations observed. In Task 1, the radiologists’ scores for teams M2 and M3 were nearly identical, consistent with their comparable SSIM values. In Task 2, team S3 received slightly higher radiologist ratings than team S2, despite similar objective metrics. The high correlation between the SSIM-based and radiologist-based rankings further confirms the consistency and reliability of the adopted evaluation procedure.} Quantitative results show that team S1/M1 achieved {statistically significant performance across all metrics in both tasks}. A representative case of the top five teams for each task is shown in Fig. \ref{fig: challenge_figure} for Task 1 and 2.

The introduction of a success-rate weight ($w$) penalizes the incomplete or partially missing reconstructions. This weighting ensures that a method’s overall score reflects not only reconstruction quality but also its robustness and completeness across all modalities, which is an essential criterion for universal reconstruction models. After applying this rule, all final submissions achieved full coverage ($w$=1.0) with no invalid outputs.

\subsection{Characteristics on effective strategies}
\subsubsection{Data processing}
From Table \ref{tab: bullet_task1} and Table \ref{tab: bullet_task2}, all the teams employed data standardization with min-max or z-score, insensitive to outliers, before input into the network training. Most teams enhanced model robustness through traditional image-domain augmentations, including flipping, rotation, and shifting. Several teams further improved generalizability by implementing domain-specific strategies such as k-space padding.

\subsubsection{Network architectures}
\begin{figure*}[t]
\centering
\includegraphics[width=\textwidth]{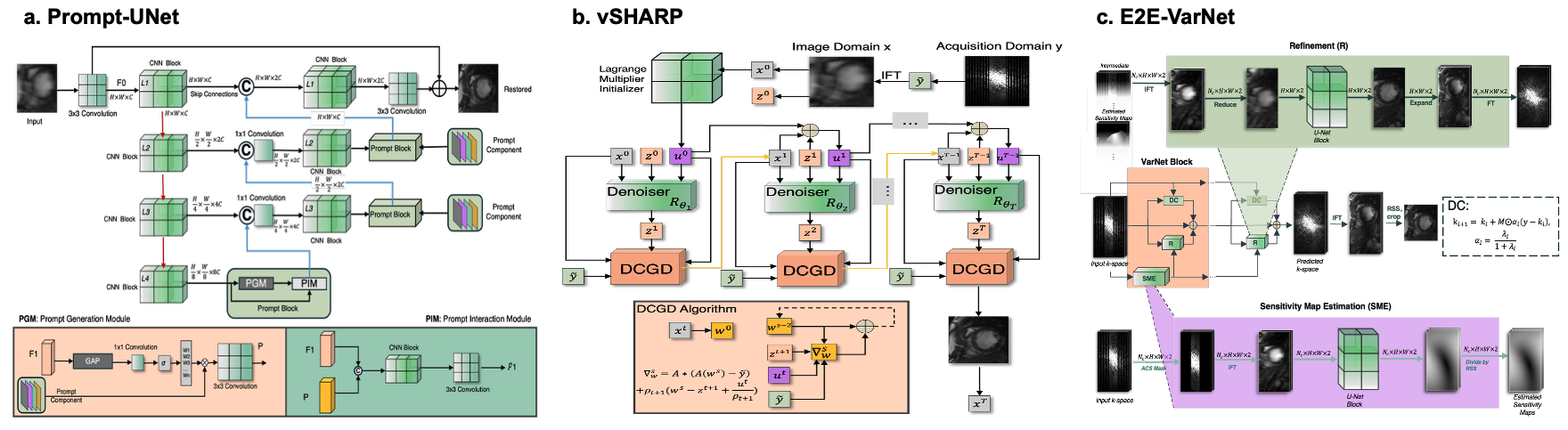}
\caption{{\textbf{Overview of top-performing team backbones.} \textbf{a. Prompt-UNet~\cite{potlapalli2023promptir}}: CNN-based architecture with prompt-driven feature modulation for MRI reconstruction.
\textbf{b. vSHARP~\cite{yiasemis2025vsharp}}: Unrolled ADMM reconstruction framework integrating learned denoisers and differentiable conjugate gradient descent (DCGD).
\textbf{c. E2E-VarNet~\cite{sriram2020end}}: End-to-end variational network combining iterative learned refinement (R), sensitivity estimation (SME), and soft data consistency (DC) for multi-coil MRI reconstruction.}
}
\label{fig: networkstructures}
\end{figure*}

E2E-VarNet~\cite{sriram2020end}, Prompt-UNet~\cite{potlapalli2023promptir} and vSharp~\cite{yiasemis2025vsharp} emerged as the dominant architectures in Figure \ref{fig: networkstructures}.

E2E-VarNet, established as the benchmark in fastMRI~\cite{zbontar2018fastmri,knoll2020fastmri}, features a coherent end-to-end learning pipeline that simultaneously optimizes three key stages: coil sensitivity map (CSM) estimation, image-domain refinement, and data consistency enforcement. The joint optimization of the spatial encoding information of CSM and the image domain in an unrolled framework creates a learnable, physics-informed reconstruction paradigm that seamlessly integrates data-driven learning with fundamental MRI physics.

Prompt-UNet~\cite{potlapalli2023promptir} was first applied for MRI reconstruction by Prompt-MR~\cite{xin2023fill}, the winner of CMRxRecon2023~\cite{lyu2024state}, leveraging prompt-based learning in a U-shape framework. Inspired by the visual prompt learning \cite{jia2022visual}, it conditioned the model on different types of inputs by injecting additional learnable parameters, allowing a single model to adapt dynamically to various tasks. 

Other network backbones were also observed across teams, including traditional U-shape architectures and their variants, which were favored for their simplicity and adaptability. Additionally, model-based approaches, such as vSHARP~\cite{yiasemis2025vsharp}, were applied by teams aiming to refine specific aspects of reconstruction. These choices showed a balance between modality-specific optimization and physical consistency for cross-scenario generalization.

\begin{figure}[t]
\includegraphics[width=\columnwidth]{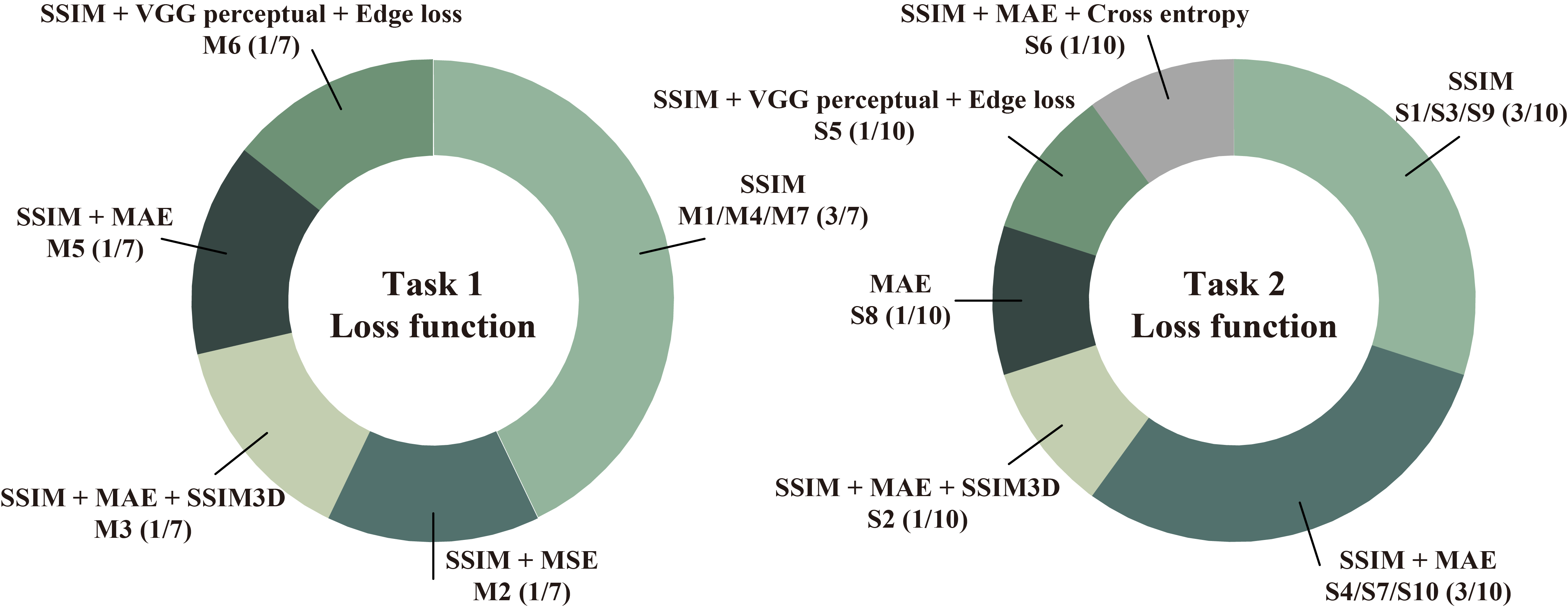}
\caption{{Loss function implemented by all participating teams in Task 1 (left) and Task 2 (right).}}
\label{fig: lossfunc}
\end{figure}

\subsubsection{Generalization ability}

To adapt to the various modalities and sampling patterns, teams implemented some information fusion strategies as shown in Table \ref{tab: bullet_task1} and Table \ref{tab: bullet_task2}.
Data balancing incorporated randomly selected acceleration factors (AFs) and sampling patterns while maintaining balance across different modalities. 
Adaptive Training approaches included curriculum learning, which gradually introduced complexity during training to optimize weight updates. Additionally, mixed precision training allowed larger model sizes while maintaining computational efficiency.
Adaptive Unrolling techniques varied across teams: team M2 implemented an unrolled discriminator, team S4 used independent network regularizers for different AFs, and team M1/S1 took additional learnable prompt embedding into the denoiser network to make it versatile. 
Multi-modality integration leveraged shared parameters and consistent loss functions to enable unified learning. 
Spatial and Temporal Attention, particularly channel-wise attention for adjacent contrast or temporal slices, was adopted. 
Group Normalization was employed to improve generalization and training stability. 
Loss function optimization included stepwise loss calculation and consistent loss functions for different modalities. 
Frequency-domain optimization optimized high-frequency details via high-pass filtering and low-frequency features for contrast.

\subsubsection{Physical measurements}
According to Table \ref{tab: bullet_task1} and Table \ref{tab: bullet_task2}, nearly all methods incorporated data consistency modules into their networks, ensuring alignment with physical measurement constraints. The primary physical modeling approaches included gradient descent (GD) and alternating direction method of multipliers (ADMM), which enforced measurement consistency during reconstruction.

\subsubsection{Loss functions}
SSIM was utilized by all teams in Task 1 and Task 2 (Fig. \ref{fig: lossfunc}), except one team, likely due to its designation as the primary ranking metric for the challenge. Mean Absolute Error (MAE) emerged as the second most frequently used metric, as it is particularly relevant to quantitative CMR tasks, where the final mapping value is a critical biomarker. Additionally, other loss functions, such as VGG perceptual loss, cross-entropy loss, and edge loss, were employed by some teams to enhance contextual and structural information in the reconstructions. 

\begin{figure*}[t]
\centering
\includegraphics[width=\textwidth]{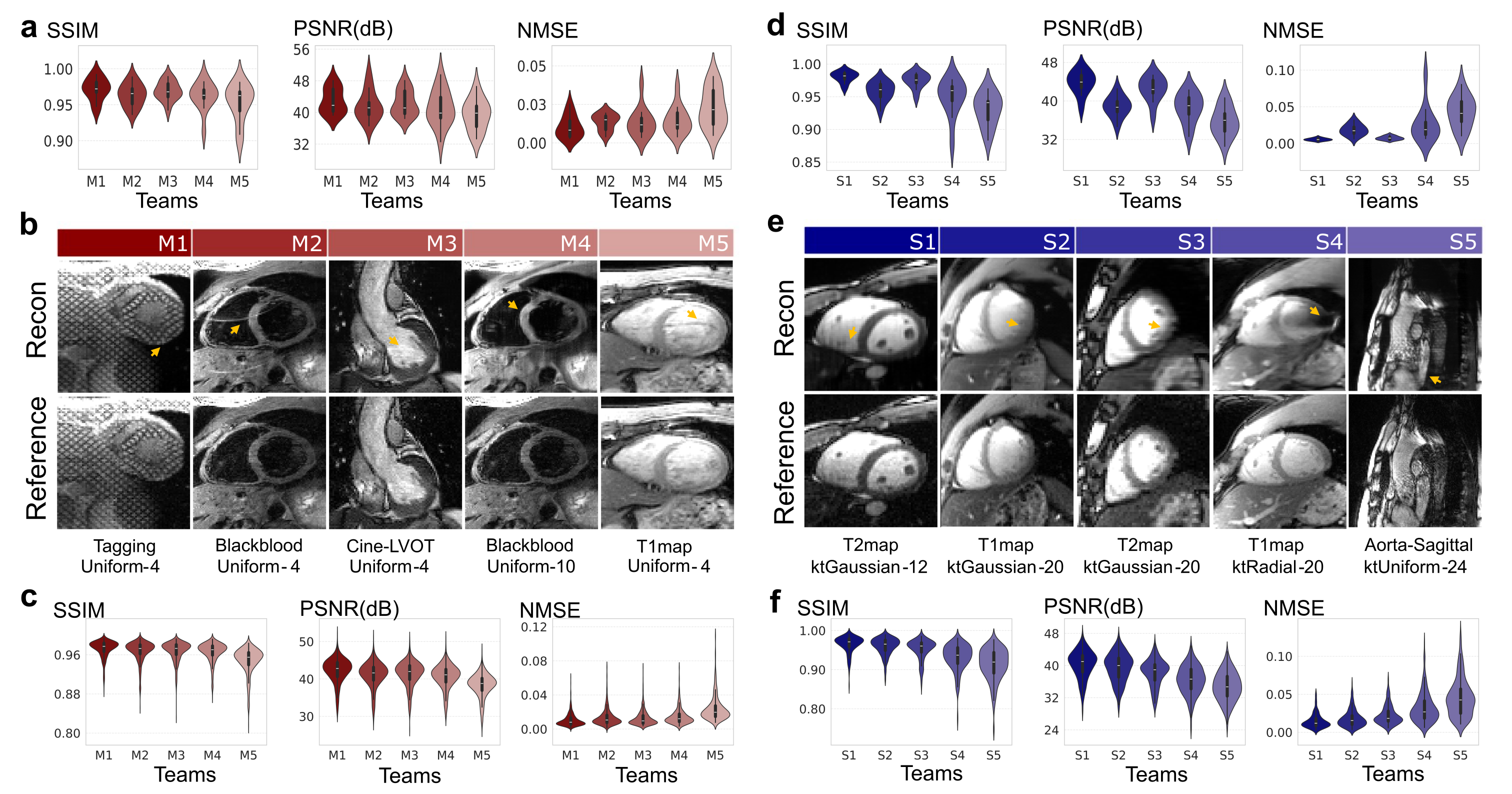}
\caption{\textbf{Failure case analysis and metric evaluation of top teams under challenging scenarios.} 
{\textbf{a.} The worst 5\% among the top five teams in Task 1.\textbf{b.} A representative case with poor image quality among the top five teams in Task 1. The undersampling patterns with the acceleration factors are labelled underneath. A fully-sampled {reference} image is put below the reconstructed image for comparison. Undesired artifacts are shown by the yellow arrow. } \textbf{c.} The SSIM, PSNR, and NMSE for the top five teams, evaluated under high acceleration factors: AF10 for Task 1. {\textbf{d.} The worst 5\% among the top five teams in Task 2. \textbf{e.} One case with poor image quality among the top five teams in Task 2. The undersampling patterns with the acceleration factors are labeled underneath. A fully-sampled reference image is put below the reconstructed image for comparison. Undesired artifacts are shown by the yellow arrow.} \textbf{f.} The SSIM, PSNR, and NMSE for the top five teams, evaluated under high acceleration factors: AF24 for Task 2.}
\label{fig: extremecases}
\end{figure*}

\begin{figure*}[t]
\centering
\includegraphics[width=\textwidth]{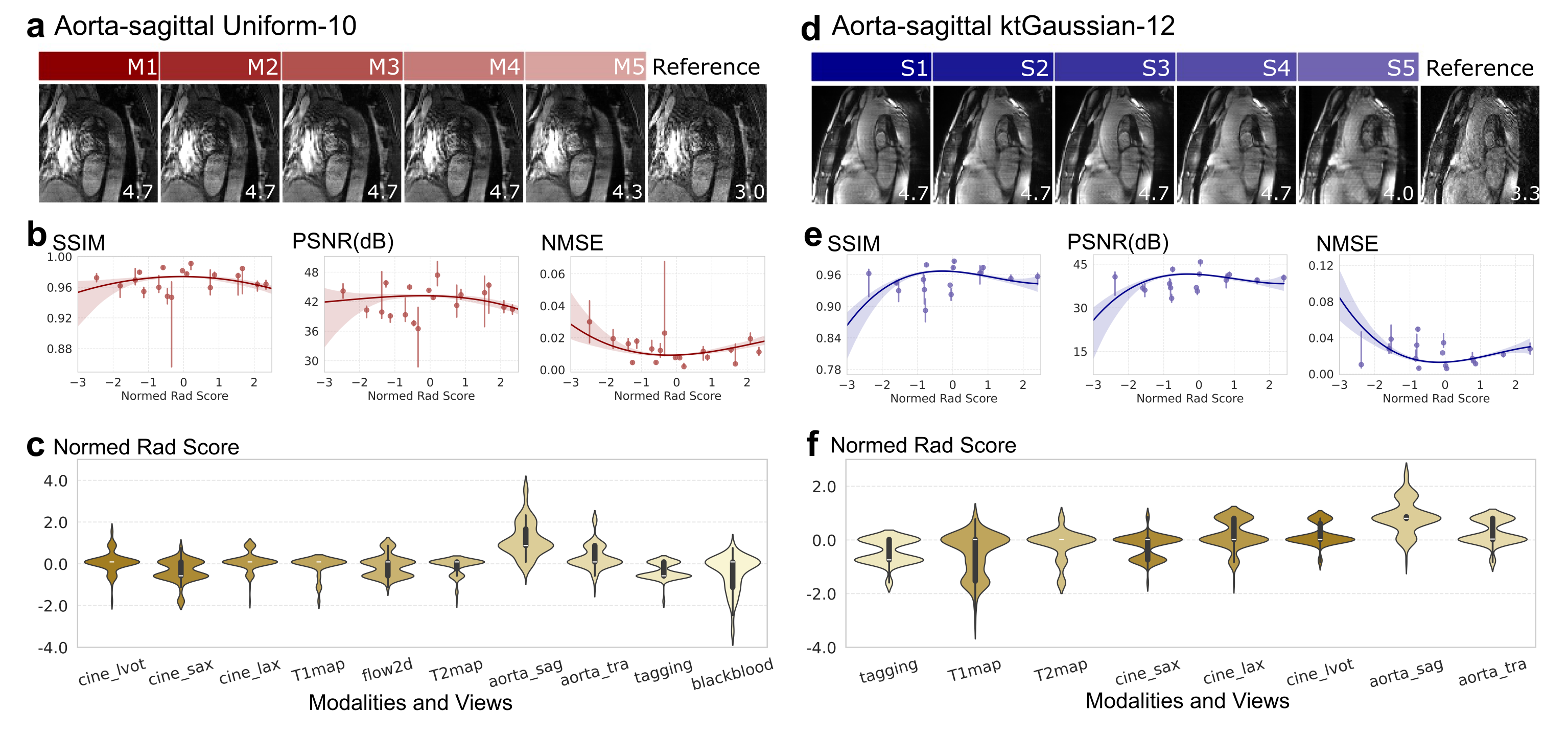}
\caption{\textbf{Objective metric and radiologist evaluation of top submissions.} \textbf{a.} One reconstructed case demonstrates superior performance compared to the {reference} image. The images showcase the outputs from the top five teams in Task 1. \textbf{b.} The SSIM, PSNR, and NMSE against the normalized radiologist score $Z_{n}$ for Task 1. \textbf{c.} The normed radiologist score $Z_{n}$ of different modalities for Task 1. \textbf{d.} One reconstructed case demonstrates superior performance compared to the {reference} image. The images showcase the outputs from the top five teams in Task 2. \textbf{e.} The SSIM, PSNR, and NMSE against the normalized radiologist score $Z_{n}$ for Task 2. \textbf{f.} The normed radiologist score $Z_{n}$ of different modalities for Task 2.}
\label{fig: subvsobj}
\end{figure*}

\subsection{Ranking stability}

Besides using the mean scores to rank the models, we analyzed the cases with the lowest 5\% performance on SSIM to assess model stability. Fig. \ref{fig: extremecases} a,d show that rankings generally remained consistent with overall performance, with a few exceptions. All other teams maintain their relative positions, demonstrating consistent performance even in challenging cases.

Additionally, we examined the worst-performing cases across different modalities for each team, as depicted in Fig. \ref{fig: extremecases} b,e. Common issues observed in the reconstructed images include oversmoothing (e.g., Team M1, S1, S3), which results in a loss of fine details; aliasing artifacts (e.g., Team M2, M5), which manifest as high-frequency noise in the images; blurriness (e.g., Team M3, S2, S5), particularly in regions with motion or complex structures; hallucination artifacts (e.g., Team S4), where spurious features appear in the image; and undesired contrast (e.g., Team M4), where the contrast does not match the reference. 

To assess the limits of extreme undersampling, we focused on analyzing the impact of high AFs: AF10 for Task 1 and AF24 for Task 2. As shown in Fig. \ref{fig: extremecases} c,f, Task 1 rankings at AF10 remained largely stable, with Teams M1 and M2 maintaining their leading positions in SSIM and PSNR, despite some variations in NMSE metrics. In Task 2 at AF24, Teams S1 and S2 preserved their performance advantages, though with more pronounced inter-team differences than in Task 1, particularly in NMSE measurements. Thus, while high acceleration factors introduced some changes in the metrics, the top teams generally maintained their rankings across both tasks. This indicates that the models are robust and capable of consistently performing under higher acceleration factors, even though small variations in the metrics are observed.

\subsection{Conflict between radiologists' ratings and objective metrics}

From the rankings for the two tasks in Table. \ref{tab: task1rank} and \ref{tab: task2rank}, we observed that the top five rankings from the radiologists sometimes conflict with the objective metrics (e.g., Team S2 and S3 in Table. \ref{tab: task1rank}).

To investigate the underlying factors that might contribute to lower-bound performance, we defined the difference between the reconstructed image from each team and the reference for each radiologist as $D_{mn}$:

\begin{equation}
D_{mn} = R_{mn} - R_{m,\text{Ref}},
\label{eq:diff_score}
\end{equation}
where $D_{mn}$ represents the difference score for the $m^{th}$ radiologist on the $n^{th}$ reconstructed output, $R_{m, Ref}$ is the radiologist's rating for the {reference} image, and $R_{mn}$ is the rating for the $n^{th}$ reconstructed output. To account for individual radiologists' variability in ratings, we applied z-score normalization, adjusting for the mean $\mu_{m}$ and standard deviation $\sigma_{m}$ of each radiologist's scores:

\begin{equation}
Z_{mn} = \frac{D_{mn} - \mu_{m}}{\sigma_{m}}.
\label{eq:z_score}
\end{equation}

This normalization transforms the scores to a zero-mean, unit-variance distribution, allowing for a more consistent comparison across radiologists. The radiologist's overall rating for the $ m^{th}$ reconstructed image was then obtained by averaging the normalized scores
$Z_{mn}$ across all radiologists, as described by Mason et al. \cite{mason2019comparison}.

Fig. \ref{fig: subvsobj} b,e illustrate the relationship between the SSIM, PSNR, NMSE metrics, and the normalized radiologist ratings $Z_{n}$ for Task 1 and Task 2. A cubic polynomial regression was performed between the radiologist score $Z_{mn}$ and the objective metrics to analyze the correlation. Since radiologist ratings were discrete, we aggregated their normalized scores using the median to minimize the influence of outliers.

As shown in Fig. \ref{fig: subvsobj} b,e, higher values on the x-axis correspond to reconstructed images that better match the reference image. Interestingly, the widely used SSIM metric, which measures image similarity to the reference, decreased as the reconstructed image performed better than the reference. This suggests that SSIM may not fully capture the perceptual quality of images that outperformed the {reference} in certain cases. Fig. \ref{fig: subvsobj} c,f present modality rankings based on normalized radiologist scores. In Task 1, black-blood imaging, an unseen modality, showed the worst performance likely due to its unique characteristics or the challenges posed by its data distribution, which may differ from the other modalities seen during training. In Task 2, the most challenging modality is tagging, as evidenced by its lower radiologist scores. {This is because the spatial tagging gradients applied during acquisition introduce periodic modulation in the myocardium, resulting in characteristic stripe and grid patterns in k-space \cite{morales2024present}. While these patterns can be accurately reconstructed under full sampling, they are particularly vulnerable to degradation under highly accelerated undersampling, posing great challenges for reconstruction methods. Nevertheless, when tagging cannot be reliably reconstructed, a feasible and efficient alternative for strain analysis is feature tracking on cine images, which is also widely used in clinical practice\cite{morales2024present}.} However, a notable trend observed in both tasks is the consistent out-performance of the aorta sagittal modality. In both Task 1 and Task 2, the reconstructed images for the aorta sagittal modality scored higher than the corresponding {reference} images, as shown by the positive radiologist scores. The aorta sagittal modality stands out as an example where the reconstructed images exceeded the original {reference} in terms of perceptual quality.

\begin{figure}[t]
\centering
\includegraphics[width=\columnwidth]{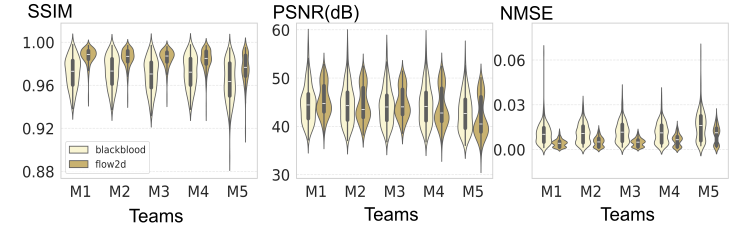}
\caption{The SSIM, PSNR and NMSE for the unseen modalities black-blood and 2D flow evaluated across the top five teams in Task 1.}
\label{fig: task1unseen}
\end{figure}

\begin{figure}[t]
\centering
\includegraphics[width=\columnwidth]{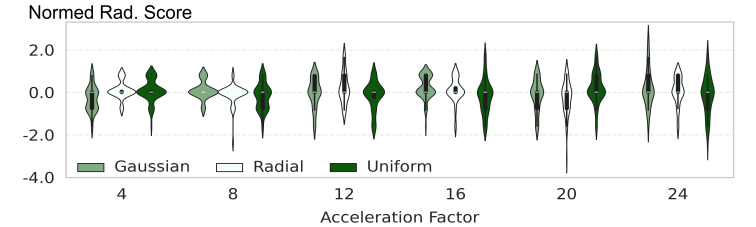}
\caption{The normed radiologist rating for different sampling patterns evaluated under different acceleration factors in Task 2.}
\label{fig: task2sampling} 
\end{figure}

\begin{figure}[t]
\centering
\includegraphics[width=\columnwidth]{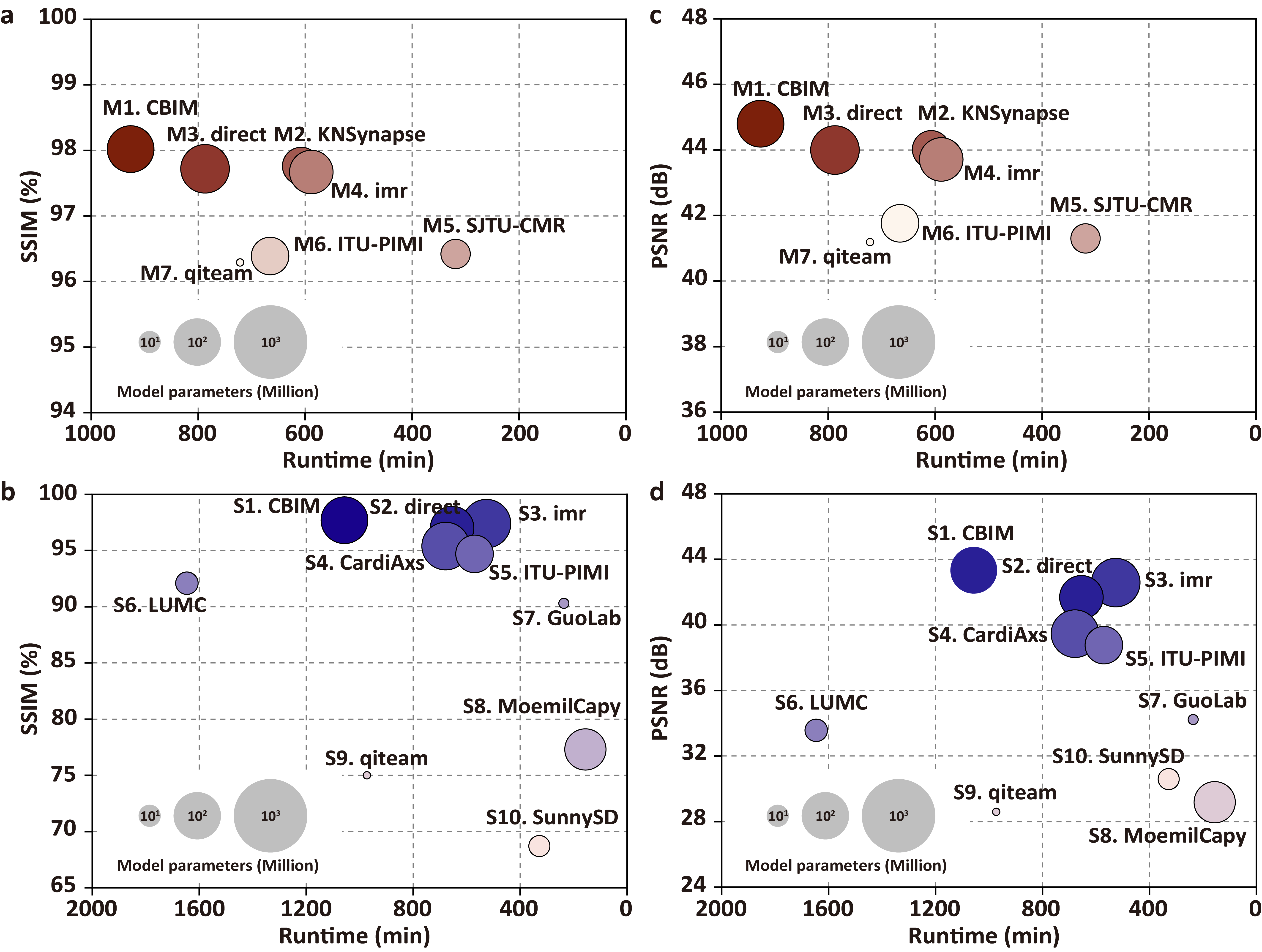}
\caption{\textbf{{Runtime} and Performance Comparison Across Ranked Teams.} Larger markers represent models with more parameters. \textbf{a.} PSNR vs. {runtime} for Task 1. \textbf{b.} SSIM vs. {runtime} for Task 1. \textbf{c.} PSNR vs. {runtime} for Task 2. \textbf{d.} SSIM vs. {runtime} for Task 2.}
\label{fig: modelpara}
\end{figure}

\subsection{Generalization to unseen modalities}
One of the core questions in Task 1 was whether a model trained on a subset of {modalities} could generalize to reconstruct entirely new modality types. We evaluated the top five teams' performance on two unseen modalities: black-blood and 2D flow. Fig. \ref{fig: task1unseen} shows that performance rankings remained largely consistent between seen and unseen modalities, indicating strong generalization capabilities. While M4 slightly outperformed M3 on 2D flow imaging in terms of SSIM and PSNR, the relative rankings for the black-blood modality remained stable, showing only minor variations in NMSE metrics. Overall, the performance on the unseen modalities is comparable to that on the seen modalities, suggesting that the models are robust and generalize well across different types of data. 

\subsection{Robustness across sampling patterns}
Task 2 pushed participants to create a model that works under a spectrum of k-space undersampling patterns and acceleration rates, from regular Cartesian masks to highly irregular pseudo-radial masks, with acceleration factors ranging from 4$\times$ up to 24$\times$. Focusing on the generalization in the sampling domain, we compared different undersampling patterns under the same acceleration factors (AFs). With increasing AFs, the behavior of the undersampling patterns becomes more evident in the SSIM values. The k-t uniform Cartesian sampling pattern showed a noticeable decrease in SSIM at higher AFs, accompanied by wider interquartile ranges, indicating increased performance variability. This degradation likely stems from insufficient k-space coverage in uniform sampling at higher AFs. In contrast, the k-t radial pattern maintained the highest SSIM values across all AFs, with consistently narrow interquartile ranges indicating stable performance. This robustness likely results from the radial pattern's structured k-space coverage. The k-t Gaussian Cartesian pattern yielded intermediate SSIM values with lower variability than k-t uniform, but remained inferior to k-t radial in consistency and overall performance. This demonstrates that while all three sampling patterns perform reasonably well under lower AFs, the k-t radial pattern stands out for its superior and stable performance at higher AFs. The increase in AF amplifies the differences between the patterns, with radial sampling emerging as the most reliable choice for maintaining high-quality reconstructions in Task 2 (Fig. \ref{fig: task2sampling}).

\subsection{Model complexity}

\begin{table*}[t]
\centering
\caption{Computational consumptions of all participating teams in Task~1, including latency and throughput analyses.}
\scalebox{0.9}{
\begin{tabular}{|l|c|c|c|c|c|c|c|c|}
\hline
\textbf{Team} & \textbf{RAM} (GB) & \textbf{GPU RAM (GB)} & \textbf{\#Params (M) } & \textbf{Runtime (h)} & \textbf{$t_\text{vol}$ (s/vol)} & \textbf{$t_\text{frame}$ (s/slice)} & \textbf{Energy (MJ)} & \textbf{Throughput (slices/s)} \\ \hline
M1. CBIM & 67.43 & 8.45 & 245 & 15.44 & 30.84 & 0.38 & 24.64 & 2.67 \\ \hline
M2. KNSynapse & 216.65 & 29.51 & 80 & 10.12 & 20.21 & 0.25 & 15.71 & 4.07 \\ \hline
M3. direct & 95.53 & 18.67 & 304 & 13.12 & 26.21 & 0.32 & 17.83 & 3.14 \\ \hline
M4. imr & 49.90 & 25.40 & 162 & 9.81 & 19.60 & 0.24 & 14.54 & 4.20 \\ \hline
M5. SJTU-CMR & 30.05 & 1.47 & 31 & 5.31 & 10.62 & 0.13 & 5.47 & 7.75 \\ \hline
M6. ITU-PIMI & 52.72 & 21.40 & 82 & 11.09 & 22.15 & 0.27 & 19.22 & 3.72 \\ \hline
M7. qiteam & 43.00 & 30.03 & 2 & 12.03 & 24.03 & 0.29 & 18.67 & 3.43 \\ \hline
\end{tabular}
}
\label{tab: modelcomput_task1}
\end{table*}

\begin{table*}[t]
\centering
\caption{Computational consumptions of all participating teams in Task~2, including latency and throughput analyses.}
\scalebox{0.9}{
\begin{tabular}{|l|c|c|c|c|c|c|c|c|}
\hline
\textbf{Team} & \textbf{RAM} (GB) & \textbf{GPU RAM (GB)} & \textbf{\#Params (M) } & \textbf{Runtime (h)} & \textbf{$t_\text{vol}$ (s/vol)} & \textbf{$t_\text{frame}$ (s/slice)} & \textbf{Energy (MJ)} & \textbf{Throughput (slices/s)} \\ \hline
S1. CBIM & 98.91 & 21.20 & 245 & 17.61 & 46.18 & 0.48 & 29.07 & 2.07 \\ \hline
S2. direct & 148.04 & 31.26 & 304 & 8.77 & 22.99 & 0.24 & 13.56 & 4.17 \\ \hline
S3. imr & 34.17 & 31.35 & 162 & 10.90 & 28.57 & 0.30 & 17.16 & 3.35 \\ \hline
S4. CardiAxs & 32.33 & 24.07 & 267 & 11.31 & 29.66 & 0.31 & 19.24 & 3.23 \\ \hline
S5. ITU-PIMI & 45.46 & 32.00 & 82 & 9.51 & 24.92 & 0.26 & 16.48 & 3.84 \\ \hline
S6. LUMC & 37.59 & 24.63 & 14 & 27.44 & 71.96 & 0.75 & 31.01 & 1.33 \\ \hline
S7. GuoLab & 248.21 & 9.79 & 3 & 3.94 & 10.32 & 0.11 & 3.81 & 9.28 \\ \hline
S8. MoemilCapy & 134.32 & 25.89 & 124 & 2.60 & 6.82 & 0.07 & 3.03 & 14.04 \\ \hline
S9. qiteam & 73.68 & 27.97 & 2 & 16.21 & 42.50 & 0.44 & 26.75 & 2.25 \\ \hline
S10. SunnySD & 124.94 & 31.35 & 12 & 5.47 & 14.34 & 0.15 & 7.47 & 6.68 \\ \hline
\end{tabular}
}
\label{tab: modelcomput_task2}
\end{table*}

Fig. \ref{fig: modelpara} visualizes the relationship between model parameters, inference time, and performance metrics, with detailed computational requirements presented in Table \ref{tab: modelcomput_task1} and Table \ref{tab: modelcomput_task2}. Unrolled optimization methods emerged as a popular baseline approach, which generally led to longer inference times. As shown in Fig. \ref{fig: modelpara}, Table \ref{tab: bullet_task1}, and Table \ref{tab: bullet_task2}, it is clear that the inclusion of unrolling plays a key role in the increased inference time. Teams that implemented unrolled optimization showed longer inference times despite having varying model sizes. In contrast, teams without unrolled optimization achieved faster inference even with larger model parameters. This suggests that the iterative nature of unrolled optimization significantly influences computational efficiency, independent of model complexity.

\section{Discussion}

The CMRxRecon2024 challenge was designed as a benchmark platform to evaluate and promote the development of {modality- and sampling-}universal learning-based reconstruction models that can be integrated into clinical applications. A key goal of the challenge was to enable these models to generalize effectively to out-of-distribution data, including unseen modalities and undersampling schemes, ensuring robust performance across diverse clinical acquisition settings. The main difference between CMRxRecon2024 and CMRxRecon2023 is that this year's challenge places a greater emphasis on the generalization ability of models. As a result, both tasks we designed aim for one-for-multiple and out-of-distribution evaluation, which are also the biggest challenges currently faced by learning-based models. By providing the code library and tutorials for various acquisition patterns, the challenge facilitated a better understanding of CMR reconstruction in clinical scenarios.

Through a detailed analysis of the results submitted to the challenge, several effective approaches emerged: 

\textbf{Improved Physical Consistency in Deep Unrolling Networks:} Integrating traditional iterative reconstruction techniques into deep unrolling networks enhances adaptability, allowing a single model to generalize across various acquisition schemes. Traditional methods~\cite{uecker2014espirit,pruessmann1999sense} estimate or acquire coil sensitivity maps (CSMs) before reconstruction, typically using a fixed calibration region in k-space. However, their accuracy deteriorates when the number of auto-calibration signal lines is limited, leading to suboptimal reconstructions.
The E2E-VarNet \cite{sriram2020end} addresses this limitation by dynamically learning CSMs as part of the network, rather than relying on precomputed maps. This is achieved through a learnable sensitivity map estimation module, which refines the CSMs iteratively based on global structural patterns in the data. To ensure stability, the maps are normalized using the Diagonal Sum-to-One constraint, preventing inconsistencies across coils. Most state-of-the-art methods follow this structure, typically updating the CSMs within the data consistency step of an unrolled network \cite{wang2024faithful}. This joint optimization of CSM estimation and image reconstruction strengthens physical consistency, ensuring that the model effectively balances data-driven learning with MRI physics.

\textbf{Prompt-based Learning with Adaptive Representations:} A universal MRI reconstruction model should be highly adaptive to diverse sampling patterns and imaging modalities, ensuring robust performance across different acquisition settings. To achieve this, many advanced methods incorporated domain-specific learnable parameters through prompt-based deep learning priors, enabling the model to dynamically adjust its feature extraction and reconstruction process based on the input characteristics. Prompt-based learning, initially popularized in natural language processing has been adapted for image restoration~\cite{potlapalli2023promptir, khattak2023maple}. The idea is to condition the model on different types of inputs by injecting additional learnable parameters as prompts, allowing a single model to adapt dynamically to different tasks. In Prompt-MR~\cite{xin2023fill}, this concept is used to condition the MRI reconstruction model based on different modalities, views, undersampling patterns, and acceleration factors. The prompts are injected at multiple levels of the encoder-decoder architecture to guide feature extraction and reconstruction. MRI reconstruction requires both global context (e.g., contrast variations across slices) and local details (e.g., preserving edge sharpness in images). By incorporating learnable parameters as prompts within the network, it can learn adaptive representations conditioned on different types of input data, making it versatile and eliminating the need for multiple separately trained models.

\textbf{Universal Models——Generalization Beyond {the Reference}:} An unexpected finding is that several methods demonstrated capabilities exceeding traditional {reference}-based reconstructions (Fig. \ref{fig: subvsobj} a,d). Although our tasks did not require participants to reconstruct images with quality exceeding the {reference}, nor did we ask them to address artifacts or noise in the acquired images, we did observe that in some results where the {reference} had flaws, the image quality achieved by the participating teams demonstrated performance surpassing that of the {reference}. While SSIM measures the similarity between the reference and reconstructed images, it is insufficient as the sole evaluation metric for assessing superior outcomes. A notable example is the aorta sagittal modality, where traditional reconstructed images exhibit inhomogeneity within the aorta. {It should be noted that, despite their clinical quality, these GRAPPA-based reference reconstructions may still contain residual artifacts or minor bias arising from the extended acquisition window and intrinsic limitations of parallel-imaging reconstruction. In this context, acquisition-related factors may also contribute to the observed differences. More complete k-space acquisitions often require longer breath-hold durations, which can increase variability due to motion artifacts, and hemodynamic turbulence in the aorta may be smoothed in sagittal reconstructions, sometimes leading to perceived improvements in image quality.} Since the input images are undersampled and the model is trained under the supervision of the reference, the universal model has the capability to enhance degraded images beyond the quality of the given {reference}. The multi-modality dataset includes high-quality cine images and comparatively lower-quality aorta sagittal images (Fig. \ref{fig: subvsobj}a,d). Consequently, the universal model, trained across these diverse modalities, was able to leverage information from higher-quality data to reconstruct images that, in some cases, exceed the reference in perceptual quality—especially when the reference image is itself suboptimal. This suggests universal models' potential in addressing broader CMR reconstruction challenges. 

\textbf{Local RAM and Computational Resources:} The multi-coil and multi-frame features of CMR data, accompanied by the huge number of model parameters, pose significant challenges for scaling universal models. Limitations in local RAM and GPU memory can hinder the development of universal models, which require substantial computational resources for training and inference. As outlined in {Table \ref{tab: bullet_task1} and Table \ref{tab: bullet_task2}}, the participating teams employed a range of hardware configurations, from high-end GPUs like the A100 to more accessible options such as the RTX3090. While these setups achieved competitive performance, the substantial computational demands of foundation models highlight the need for innovative strategies to reduce resource requirements. Meanwhile, models with a larger number of parameters tend to exhibit better universal reconstruction capabilities, implying that the continuous acquisition of larger-scale datasets is likely to be pivotal in driving the ongoing enhancement of performance.

\textbf{Evaluation Metrics——Inconsistency Between Radiologists' Rating and Objective Metrics:} We examine a potential conflict between radiologists' rating and SSIM as evaluation metrics, as shown in Table \ref{tab: task1rank} and \ref{tab: task2rank}. SSIM, widely used in reconstruction challenges \cite{zbontar2018fastmri, lyu2024state}, emphasizes pixel-wise structural similarity. However, it does not always align with radiological interpretations that are crucial for clinical decision-making \cite{mason2019comparison}. This disparity underscores the need for evaluation metrics that balance structural fidelity with clinical relevance. {Regarding the reader study, we acknowledge that restricting evaluation to the top five teams may introduce selection bias by excluding lower-ranked outputs. This design choice, however, was primarily driven by the practical constraints of the radiologists’ workload. Each radiologist needs to evaluate more than 1,000 images per task, encompassing all modalities, sampling patterns, and anatomical views. This extensive reading already represented a substantial commitment of time and effort. Under these circumstances, limiting the reader study to the top-performing teams was a balanced compromise that ensured full modality coverage and statistical robustness while maintaining feasibility. Ideally, future editions could extend the evaluation to all teams if resources allow, but the current design provides a rigorous and representative analysis within practical constraints. Moreover, this setting was announced at the beginning of the challenge, and therefore we need to follow this predefined scheme for the final ranking.
Additional inter-reader reliability was quantitatively assessed using ICC (two-way mixed). Based on reader-wise z-normalized scores, the agreement reached a moderate level (Task 1: ICC(3,k) = 0.353; Task 2: ICC(3,k) = 0.379), suggesting that the three radiologists exhibited similar ranking tendencies but differed in their individual scoring scales. An important factor that limit the scoring consistency is that all datasets have undergone strict quality control before release, and images with clearly poor reconstruction quality were excluded. As a result, most of the evaluated images were of high quality, and the remaining perceptual differences were subtle, which further limited the inter-reader variance. Another important factor is the intrinsic subjectivity of qualitative image assessment, as radiologists may focus on different perceptual attributes such as artifact suppression, texture naturalness, or edge sharpness when forming their overall judgments. To address these challenges, future reader studies will adopt a structured multi-dimensional scoring protocol that separates evaluations of artifacts, contrast, texture, and anatomical delineation, supported by standardized scoring guidelines and representative examples. This design is expected to improve inter-reader consistency, enhance interpretability, and strengthen the clinical relevance of the evaluation.}

{As part of the CMRxRecon series, we have been building the dataset and challenge gradually and strategically. Following last year's single-modality reconstruction task on cine and mapping, the present CMRxRecon2024 phase focuses on developing and validating a modality- and sampling-universal reconstruction framework across multiple sequences, including tagging, 2D flow, black-blood, LVOT, aorta-tra and aorta-sag. In this phase, our key priorities are to establish a well-controlled benchmark using 3T data from healthy volunteers acquired on a single vendor and within a single centre, ensuring technical consistency and facilitating cross-modality evaluation. Building on this foundation, subsequent phases of the CMRxRecon series will progressively extend toward multi-centre, multi-vendor datasets that include both healthy participants and patients with diverse cardiovascular diseases such as hypertrophic and dilated cardiomyopathy, myocardial infarction, coronary artery disease and arrhythmias. Data acquisition across scanners from GE, Philips, Siemens and United Imaging, covering field strengths beyond 3T and sequences such as TrueFISP (cine and tagging), FLASH (mapping and dark-blood) and TSE (T2-weighted imaging), will enable vendor-independent benchmarking and field-strength robustness analyses. We also plan to include electrocardiogram (ECG) signals to enable more accurate synchronization of image acquisition with cardiac phases. This will allow more reliable quantification of dynamic cardiac function, blood flow parameters, and myocardial tissue characterization. In parallel, future work will extend the reconstruction framework from magnitude-focused to phase-resolved imaging, enabling the recovery of velocity-encoded and flow-sensitive information. This expansion will support phase-contrast MRI applications for quantitative assessment of hemodynamic parameters and velocity fields under different VENC settings, advancing universal reconstruction toward comprehensive, physiology-informed cardiac imaging. In addition, quantitative domain-shift analyses (e.g., cross-vendor or acceleration-factor–stratified error) are planned within the upcoming multi-centre extensions to evaluate model generalizability. Although the current reference images are reconstructed using clinically adopted 2$\times$ GRAPPA to balance scan efficiency and image quality, such reconstructions may still contain residual artifacts or bias due to extended acquisition windows and the intrinsic limitations of parallel imaging, particularly in challenging regions such as the aorta where complex flow patterns can amplify inconsistencies. Systematic analyses will help distinguish genuine reconstruction improvements from GRAPPA-related differences, ensuring that benchmark performance reflects true methodological advances. Finally, while CMRxRecon2023 relied primarily on objective image-quality metrics, the current challenge introduces expert radiologists’ evaluations to complement quantitative measures. In this phase, we prioritize reproducibility and interpretability across modalities, while the challenge will continue to evolve with the inclusion of clinically meaningful metrics such as segmentation accuracy, functional indices, and physiologically relevant parameters (e.g., biventricular volumes, T1/T2-mapping accuracy, strain and aortic flow). Through this continuous and open evaluation design, we aim to progressively bridge quantitative assessment with clinical interpretation and foster the development of reconstruction methods aligned with diagnostic needs.}

Technical advances in MRI encoding and reconstruction demonstrated by this challenge offer potential clinical benefits. Improved reconstruction techniques reliably generate high-quality images from undersampled data, substantially enhancing visual image assessment through increased image contrast, reduced noise, and clearer anatomical definition. These improvements empower radiologists with greater diagnostic confidence, reduce interpretation variability, and significantly decrease cognitive workload. Shorter scan times directly enhance patient comfort, reduce motion artifacts and broaden access to comprehensive imaging protocols—facilitating earlier disease detection and precision treatment planning. Furthermore, the production of large, high-quality imaging datasets through these accelerated reconstruction methods strongly supports advanced downstream analyses such as automated segmentation, radiomics, and predictive modeling. Ultimately, the clinical value of the CMRxRecon2024 challenge lies in translating cutting-edge AI research into routine clinical practice, improving patient outcomes, and advancing personalized cardiovascular healthcare.

\section{Acknowledgment}
We also want to thank Meng Ye and Leon Axel from Team CBIM, Yajing Zhang from Team CMRxRecon2024-qiteam, Fumin Guo from Team GuoLab, Calder Sheagren from Team SunnySD, Lijun Zhang and Yi Chen from Team imr, and Chinmay Rao from Team LUMC for their efforts in the challenge.

\bibliographystyle{IEEEtran}
\bibliography{references.bib}
\end{document}